\documentclass[manuscript,screen,nonacm]{acmart}

\AtBeginDocument{%
  \providecommand\BibTeX{{%
    \normalfont B\kern-0.5em{\scshape i\kern-0.25em b}\kern-0.8em\TeX}}}

\setcopyright{acmcopyright}
\copyrightyear{2018}
\acmYear{2020}
\acmDOI{10.1145/1122445.1122456}

\acmJournal{TDS}
\acmVolume{1}
\acmNumber{1}
\acmArticle{1}
\acmMonth{1}

\usepackage{caption}
\usepackage{subfigure}
\usepackage{multirow}
\begin{document}

\title{AI Marker-based Large-scale AI Literature Mining}

\author{Rujing Yao}
\authornote{Both authors contributed equally to this research.}
\email{rjyao@tju.edu.cn}
\author{Yingchun Ye}
\authornotemark[1]
\email{ycye@tju.edu.cn}
\affiliation{%
  \institution{Center for Applied Mathematics, Tianjin University}
  \streetaddress{92 Weijin Road}
  \city{Nankai Qu}
  \state{Tianjin Shi}
  \postcode{300072}
  \country{China}
}

\author{Ji Zhang}
\email{zhangji77@gmail.com}
\affiliation{%
  \institution{Insitute of AI, Zhejiang Lab}
  \streetaddress{1818 Wenyi West Road}
  \city{Yuhang Qu, Hangzhou Shi}
  \state{Zhejiang Sheng}
  \country{China}
}

\author{Shuxiao Li}
\email{shuxiao.li@ia.ac.cn}
\affiliation{%
  \institution{Institute of Automation, Chinese Academy of Sciences}
  \streetaddress{95 Zhongguancun East Road}
  \city{Haidian Qu}
  \state{Beijing Shi}
  \postcode{100190}
  \country{China}
}

\author{Ou Wu}
\email{wuou@tju.edu.cn}
\authornote{Corresponding author}
\affiliation{%
  \institution{Center for Applied Mathematics, Tianjin University}
  \streetaddress{92 Weijin Road}
  \city{Nankai Qu}
  \state{Tianjin Shi}
  \postcode{300072}
  \country{China}
}

\renewcommand{\shortauthors}{Rujing Yao and Yingchun Ye, et al.}

\begin{abstract}
The knowledge contained in academic literature is interesting to mine. Inspired by the idea of molecular markers tracing in the field of biochemistry, three named entities, namely, methods, datasets and metrics are used as AI markers for AI literature. These entities can be used to trace the research process described in the bodies of papers, which opens up new perspectives for seeking and mining more valuable academic information. Firstly, the entity extraction model is used in this study to extract AI markers from large-scale AI literature. Secondly, original papers are traced for AI markers. Statistical and propagation analysis are performed based on tracing results. Finally, the co-occurrences of AI markers are used to achieve clustering. The evolution within method clusters and the influencing relationships amongst different research scene clusters are explored. The above-mentioned mining based on AI markers yields many meaningful discoveries. For example, the propagation of effective methods on the datasets is rapidly increasing with the development of time; effective methods proposed by China in recent years have increasing influence on other countries, whilst France is the opposite. Saliency detection, a classic computer vision research scene, is the least likely to be affected by other research scenes.
\end{abstract}



\maketitle

\section{Introduction and Related work}
The exploration of academic literature can help researchers quickly and accurately seek research information and understand research trends. At present, most literature research heavily relies on the metadata of papers, including authors, titles, citations and so on. The number of authors is used by Sahu et al. to evaluate the effect on the quality of the literature \cite{r1}. Wang et al. released a ranking list of highly cited scholars in the AI field by statistics on the number of citations\footnote{\url{https://www.acemap.info/ranking}}. Citation count is used to estimate future citations for literature by Yan et al. \cite{r2}. Li et al. used a knowledge graph derived from literature metadata to compare entity similarities (papers, authors and journals) in the embedding space \cite{r3}. Tang et al. studied the evolution trend of the AI field on the basis of keywords and the countries of authors \cite{r4}. Numerous studies have analysed literature based on authors, keywords, citations and so on \cite{r5,r6,r7,r8,r9}.

Some researchers focus on abstracts of papers because metadata contains limited semantic information. An abstract is a high-level summary of the content of a paper. Topic model is the primary analytical tool \cite{r10,r11,r12,r13,r14,r15}. Iqbal et al. used Latent Dirichlet Allocation (LDA) to model important topics in COMST and TON \cite{r16}. Tang et al. used Author-Conference-Topic model to construct academic social networks \cite{r17}. Moreover, hot research topics in AI are sought by Tang et al\footnote{\url{https://www.aminer.cn/ai2000}}. However, a problem of topic granularity inconsistency occurred when the topic models are used on abstracts. For example, in the seeking of top 10 hot research topics conducted by Tang et al., the granularities of the topics are quite inconsistent. For instance, Neural Network, Convolutional Neural Network and Machine Learning, which are in distinct research levels, are listed amongst the top 10 topics.

Abstracts mainly contain conclusive information and lack information reflecting the research process. The bodies of the papers contain the specific process of research. However, little research has focused on paper bodies. One of the main reasons for this phenomenon is that the body of a paper contains thousands of words. Applying the existing topic models may cause problems, such as some non-topic frequent words in the texts with low topic relevance may be used as topic words.

\begin{figure}[htbp]
\setlength{\belowcaptionskip}{-0.01cm}
\centering
\subfigure[The isotope of oxygen ${}^{18}O$ is used by Samuel Ruben and Martin Kamen to mark ${H_2}O$ and $C{O_2}$ respectively, and to track the source of ${O_2}$ in photosynthesis.]{

\begin{minipage}[b]{0.5\textwidth}

\includegraphics[width=1\textwidth]{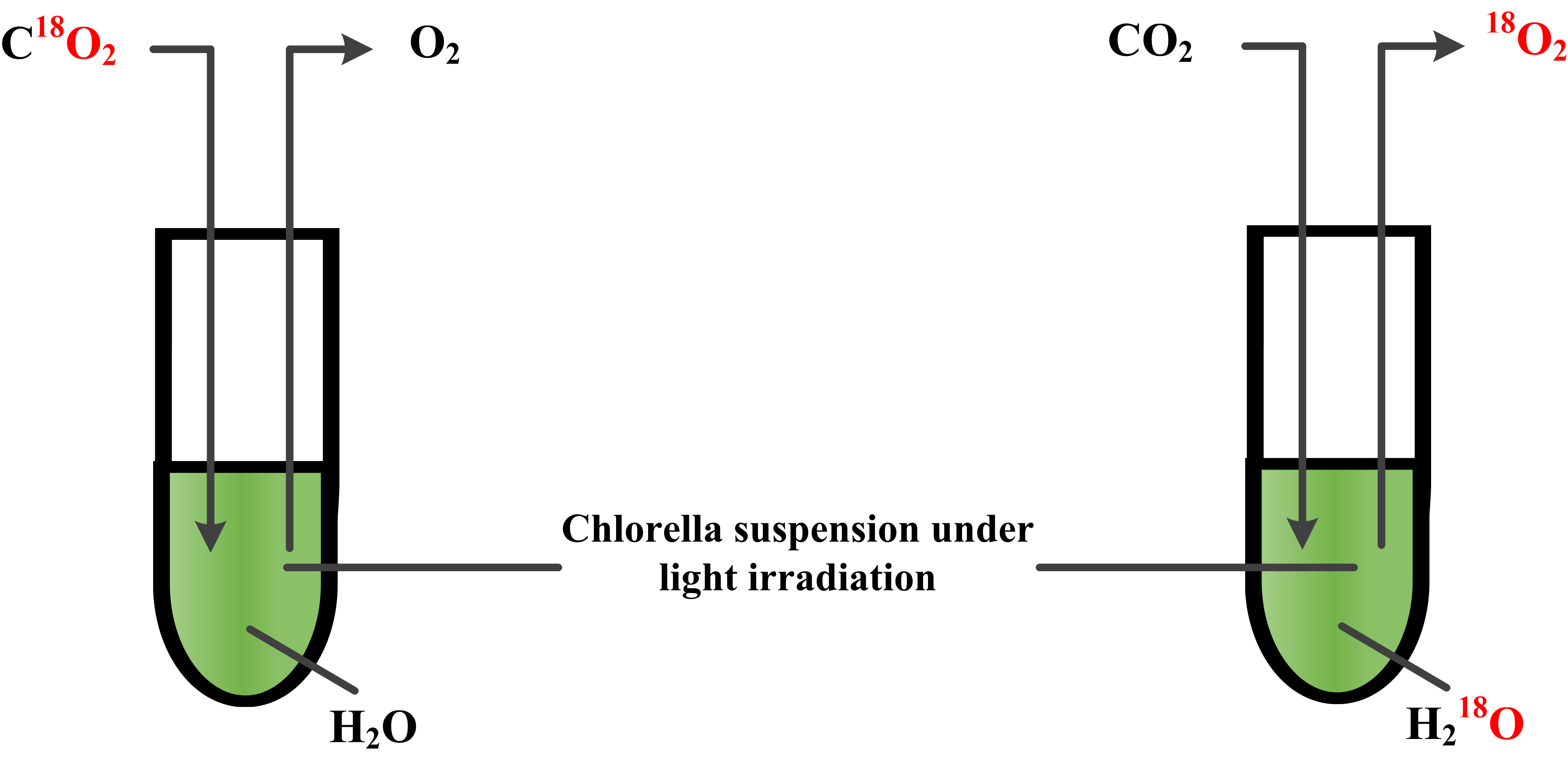}
\label{fig:fig1a}
\end{minipage}

}

\subfigure[When the AI markers are proposed or cited by different literature, the traces in the specific research process are formed. Therefore, the AI markers can play the same role as molecular markers in the mining of the characteristics and regularities of the literature. ]{
\begin{minipage}[b]{0.5\textwidth}
\includegraphics[width=1\textwidth]{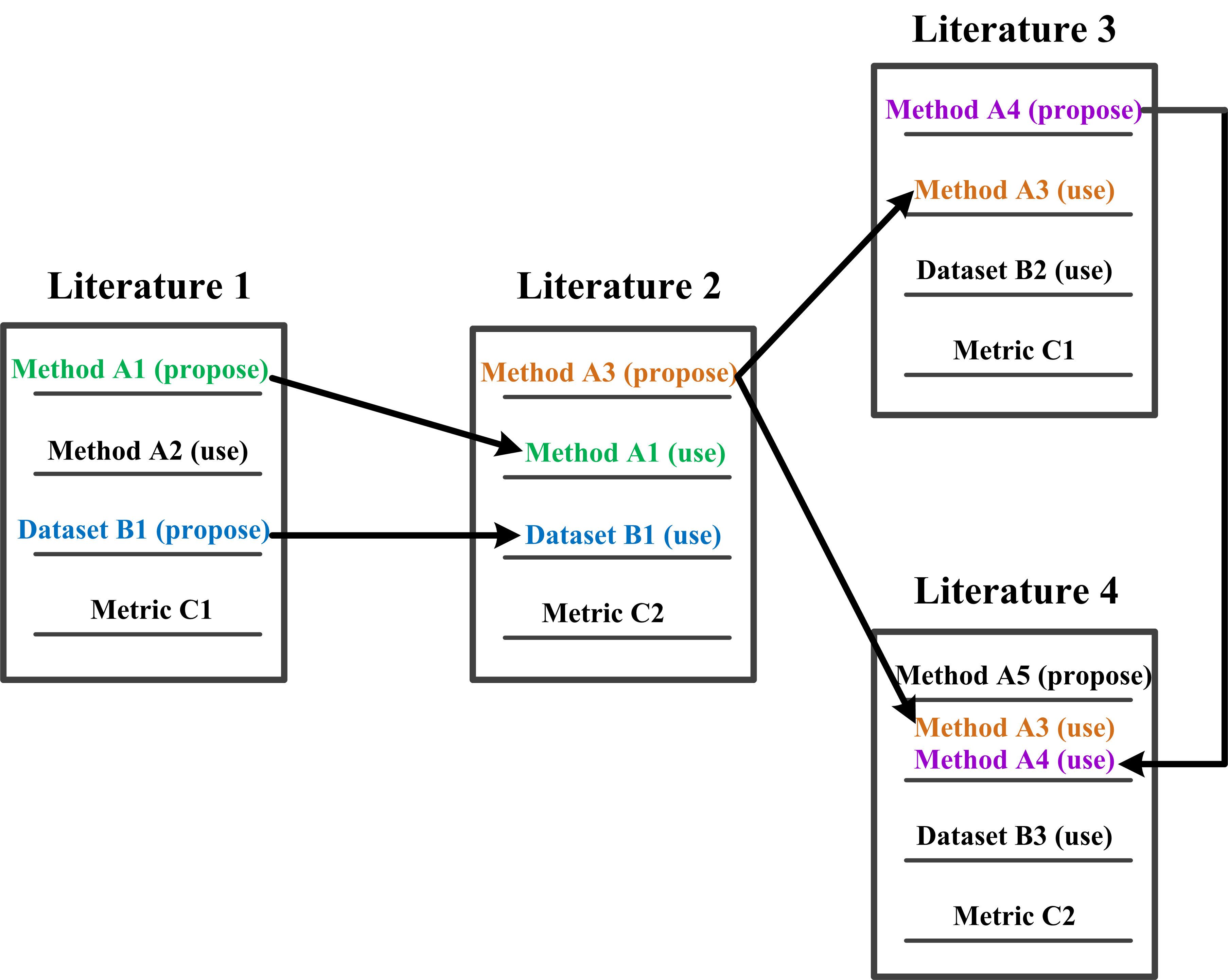}
\label{fig:fig1b}
\end{minipage}
}
\caption{Analogy diagram of the molecular and AI markers.} \label{fig:fig1}
\end{figure}

In the biological field, molecular markers are often used to track the changes in the substances and cells during the reaction, to obtain reaction characteristics and regularities \cite{r18,r19}. Inspired by this, we observe that methods, datasets and metrics can play the same role as molecular markers in AI literature mining. The methods, datasets and metrics, which are in the same granularity by large, in the AI literature are used as AI markers. These factors can be leveraged to trace the information reflecting the research process in the paper bodies rather than metadata and abstracts. Fig. \ref{fig:fig1} describes the similarity with AI markers and molecular markers. The AI maker-based mining supplements conventional metadata/abstract-based literature mining.

In our work, an entity extraction model is proposed to firstly extract AI markers from large-scale AI literature. Effective markers (methods and datasets) are statistically analysed. Secondly, we trace the original papers for effective methods and datasets. Statistical analysis for the original papers and method propagation in different countries and datasets are conducted. Finally, clustering is performed to obtain method clusters and research scene clusters according to the co-occurrences of AI markers. Method roadmap generation and influence analysis are conducted on the basis of the two types of clusters.

The following main observations and conclusions are obtained:

1) The annual development of the AI field is obtained on the basis of the extracted effective AI markers (methods and datasets). For example, KITTI, a classic dataset in the field of driverless, is amongst the top 10 datasets in 2017, thus indicating that driverless is a major research topic in 2017;

2) Based on the original paper tracing for AI markers, the number of effective methods proposed by Singapore, Israel and Switzerland is relatively large. From the relationships between effective methods and datasets, the propagation of the effective methods on the datasets is rapidly increasing with the development of time. Based on the propagation analysis of the effective methods amongst countries, the effective methods proposed by China have an increasing influence on other countries, whilst France is the opposite;

3) Method roadmaps are constructed on the basis of the method clusters and associated datasets, which can show the evolution in method clusters. In research scene clusters, saliency detection, a classic computer vision research scene, is the least likely to be affected by other research scenes.
\section{Data}
A large amount of AI literature is necessary for our study. Firstly, the section introduces the literature data we collected. In addition, two machine learning models are used during the research. Therefore, the section also introduces the training data of these two models.

\subsection{Collected literature data}

A total of 122,446 papers published from 2005 to 2019 were collected by using the list of AI journals and conferences in China Computer Federation (CCF)\footnote{CCF compiled a list of AI journals and conferences with different ranks. See \url{https://www.ccf.org.cn/Academic\_Evaluation/By\_category/} for details.} ranks (Tier-A, Tier-B, and Tier-C). The GROBID\footnote{\url{https://grobid.readthedocs.io/en/latest/}} is utilised to convert PDF format papers into XML format. The data is obtained by extracting certain pieces of information, such as titles, countries, bodies and references, from the papers in XML format. To facilitate reading, the collected data is called CCF corpus.

\subsection{Training data for the chapter classification}

The main body of an AI paper generally includes four chapters: introduction, methodology, experiment and conclusion. The roles of AI markers in different parts vary. This study introduces a chapter classification strategy to divide the body of AI literature into the above-mentioned four parts.

2000 papers in the CCF corpus are randomly selected to train the chapter classifier. Ten graduate students major in AI are recruited to label the 2000 papers with 63,110 paragraphs. This data corpus is called TCCdata and is used to construct a BiLSTM \cite{r20} classifier for chapter classification. The numbers of chapters and the associated numbers of paragraphs for each chapter in TCCdata are shown in Table~\ref{tab:tab1}.

\begin{table*}
  \caption{Numbers of chapters and paragraphs in the TCCdata}
  \label{tab:tab1}
  \begin{tabular}{ccc}
    \toprule
     & Chapters & Paragraphs\\
    \midrule
    "Introduction"&2944 & 24674\\
    "Methodology"&985 & 14135\\
    "Experiment"&1948 & 18185\\
    "Conclusion"&2397 & 6116\\
    \bottomrule
  \end{tabular}
\end{table*}
\subsection{Training data for the AI marker extraction}

To learn an AI marker extraction model, 1000 papers from the CCF corpus are randomly selected. The "methodology" and "experiment" chapters of the 1000 papers are divided into sentences according to punctuation. Ten graduate students major in AI are recruited to label the sentences. The BIO labelling strategy is adopted for the three AI markers, namely, methods, datasets and metrics. The set of methods, datasets and metrics compiled by JiqiZhixin is used as annotation references\footnote{\url{https://www.jiqizhixin.com/sota}}. Finally, 10,410 labelled sentences are obtained and are called TMEdata.

During the training of the extraction model, TMEdata is divided into training, validation and testing sets according to the ratio of 7.5:1.5:1. The details are shown in Table~\ref{tab:tab2}.

\begin{table}[h]
\caption{Numbers of AI markers in the TMEdata}
\label{tab:tab2}
\begin{tabular}{@{}ccccc@{}}
\toprule
\multicolumn{2}{c}{\multirow{2}{*}{}}               & \multicolumn{3}{c}{AI markers} \\\cmidrule(l){3-5}
\multicolumn{2}{c}{}                                & Method   & Dataset   & Metric  \\
\midrule
\multirow{2}{*}{Training}   & Without deduplication & 4620     & 2492      & 1012    \\
                            & Deduplication         & 3821     & 1534      & 388     \\
\multirow{2}{*}{Validation} & Without deduplication & 793      & 519       & 196     \\
                            & Deduplication         & 621      & 431       & 137     \\
\multirow{2}{*}{Testing}    & Without deduplication & 554      & 292       & 154     \\
                            & Deduplication         & 482      & 259       & 67      \\
\bottomrule
\end{tabular}
\end{table}

\section{Methodology}

The section introduces the specific methods involved in the research, including chapter classification, AI markers extraction and normalisation, original paper tracing for AI markers, clustering of methods and research scenes, roadmap generation for method clusters and influencing degrees for research scene clusters.

\subsection{Chapter classification}
In the body of an AI paper, the AI markers located in the "methodology" and "experiment" chapters play a substantial role in the paper. Accordingly, only the AI markers of the "methodology" and "experiment" chapters are extracted. Simple rule strategies are difficult to use to accurately classify the chapters of the AI literature due to the diversity of the structure of the AI literature. Therefore, the chapter classification strategy that combines the BiLSTM classifier and rules is adopted.
\subsubsection{Proposed classification strategy}

\begin{figure}[htbp]
\setlength{\belowcaptionskip}{-0.3cm}
    \centering
    \includegraphics[width=\linewidth]{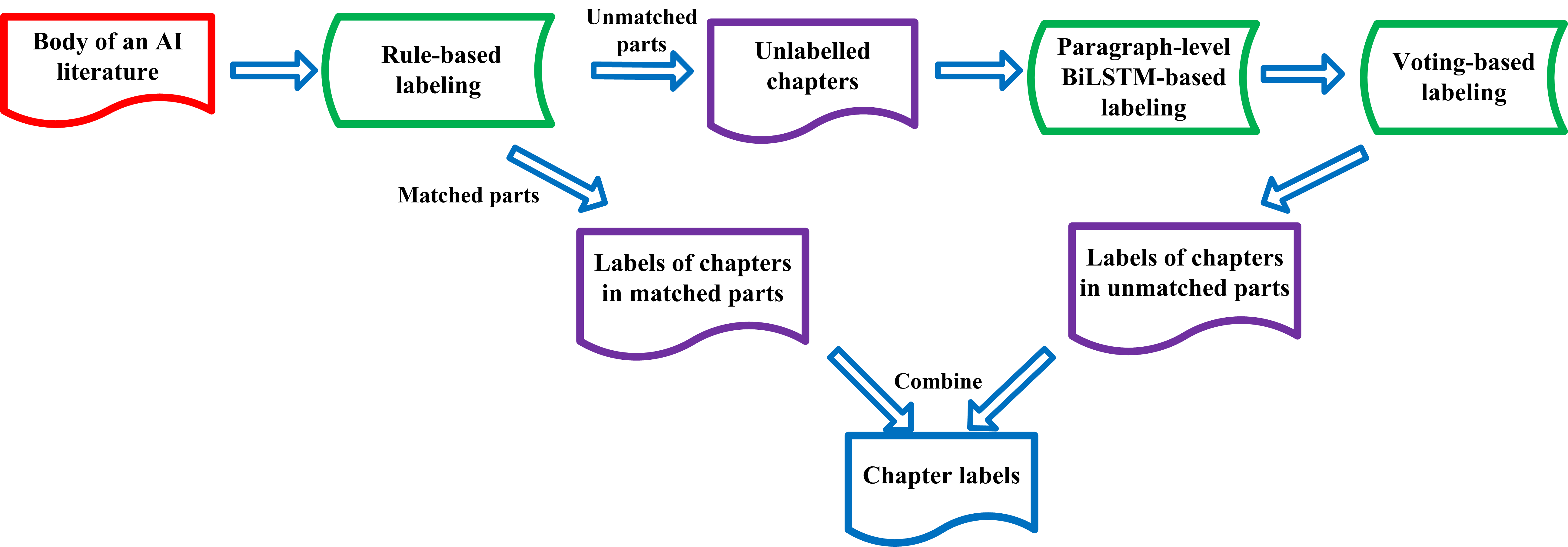}
    \caption{Overall process of chapter classification.}
    \label{fig:fig2}
\end{figure}

The overall of our strategy is shown in Fig. \ref{fig:fig2}. Rules (e.g. keywords and orders) are firstly used to label the chapters. In well-matched chapters, the chapter labels are outputted. In unmatched chapters, the paragraphs under the chapters are inputted into the paragraph-level BiLSTM classifier trained based on the TCCdata for prediction. Next, the paragraph-level predicted labels in the same chapter are voted to generate the final label. Finally, rule-based and BiLSTM-based results are combined to obtain the final chapter labels of the whole body.

The conventional one-layer BiLSTM architecture is adopted. The maximum sentence length is 200, the dimension of the word vector is 200, the hidden dimension is 256, and the batch size is 64. Cross entropy is used as the loss function. TCCdata is used as the training data.

\subsubsection{Evaluation results}
The TCCData is divided into training, validation and testing sets according to the ratio of 8:1:1. Three methods, namely, rule-based, BiLSTM-based and combining-based, are evaluated. The results show that the accuracy is 0.793 by only using rule matching. The accuracy is 0.792 by only using paragraph-level BiLSTM trained on TCCData. The accuracy reached 0.928 by using the combination.

\subsection{AI marker extraction and normalisation}
The extraction and normalisation of AI markers is challenging. Given that a large number of AI literature emerge every year, the number of new AI markers continues to increase, and the forms vary. Some common words may also be used as datasets, such as ``DROP'' in \cite{r21}. No prescribed standard is imposed for the naming of AI markers. Furthermore, some AI markers have ambiguity issues. For example, CNN can represent the Cable News Network dataset and the Convolutional Neural Networks method. LDA can represent Latent Dirichlet Allocation and Linear Discriminant Analysis.

\subsubsection{AI marker extraction model}
\begin{figure}[htbp]
\setlength{\belowcaptionskip}{-0.3cm}
    \centering
    \includegraphics[width=\linewidth]{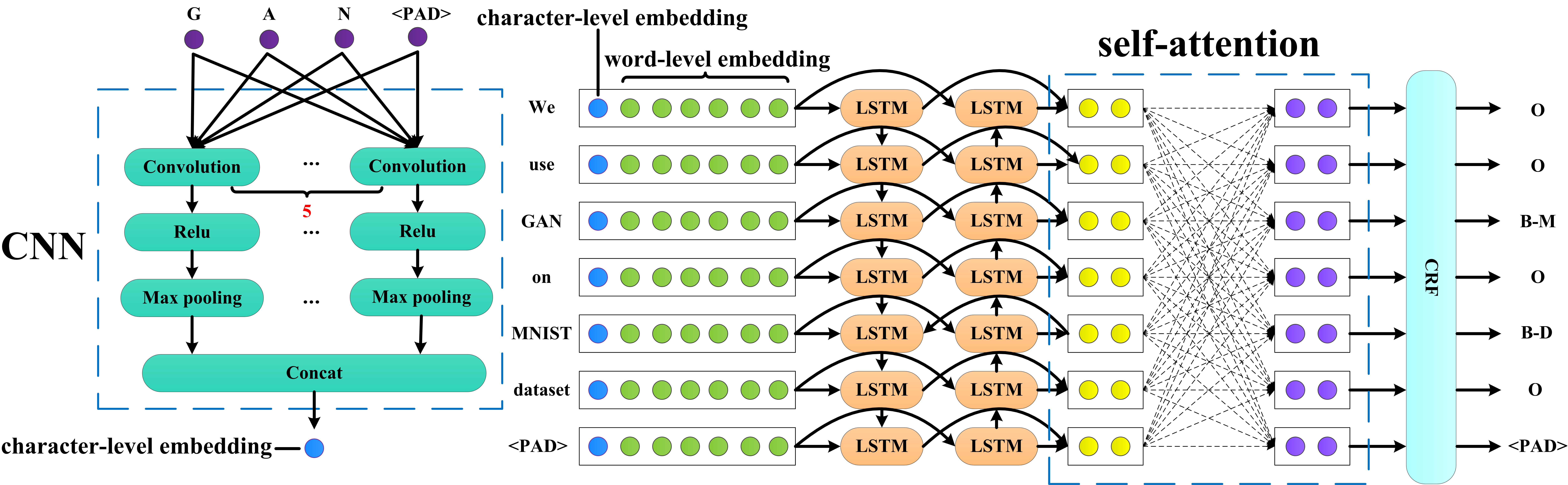}
    \caption{Structure of the AI marker extraction model.}
    \label{fig:fig3}
\end{figure}

AI marker extraction is a typical named entity recognition problem. The network structure of the adopted AI marker extraction model is based on the classical CNN+BiLSTM+CRF framework \cite{r22} with minor improvements (Fig. \ref{fig:fig3}).

In an input sentence $\left\{ {{{\rm{w}}_{\rm{1}}},{\rm{ }}{{\rm{w}}_{\rm{2}}},{\rm{ }}...,{\rm{ }}{{\rm{w}}_{\rm{n}}}} \right\}$, ${w_i}$ represents the $i$th word. Firstly, the character-level embeddings of every word are obtained through the CNN network. Secondly, the word-level embedding of every word is obtained through the Glove \cite{r23}. The above-mentioned two embeddings are co-created to be fed into Bi-LSTM. Self-attention \cite{r24} is used to calculate the association amongst words. Finally, the hidden vector obtained through self-attention is sent to the CRF \cite{r25} to obtain the label sequence $y$ of every word. ${\rm{y}}\in\left\{ {{\rm{B}}-{\rm{M}}, {\rm{I}}-{\rm{M}}, {\rm{B}}-{\rm{D}}, {\rm{I}}-{\rm{D}}, {\rm{B}}-{\rm{C}}, {\rm{I}}-{\rm{C}}, {\rm{O}}, {\rm{padding}}}\right\}$, corresponding to methods, datasets, metrics and others.

\subsubsection{Experimental settings}

The model parameters are set as follows. The maximum length is 100, the maximum word length is 50, and the batch size is 16. The character-level CNN network uses five parallel 3-dimensional convolution-activation-max pooling. Each of the five convolutions uses 10 3-dimensional convolution kernels (10 1*1*50, 1*2*50, 1*3*50, 1* 4*50 and 1*5*50). Relu is used as the activation function. Finally, the results obtained by five convolutions are spliced to obtain a 50-dimensional character-level embedding for every word. One-layer Bi-LSTM is selected, the hidden dimension of Bi-LSTM is 200, and the hidden dimension of self-attention is 400.

\subsubsection{Evaluation results}

The raw and the corresponding lowercase samples are used to train the model. During the test, the test samples (1040 sentences) and the corresponding 1040 lowercase samples are tested. The evaluation results of the AI marker extraction model are shown in Table~\ref{tab:tab3}.

Table~\ref{tab:tab3} illustrates that our model outperforms the traditional CNN+BiLSTM+CRF model. After combining with rules, such as black and white lists, the F1 of our model is 0.864, the recall is 0.876, and the precision is 0.853.

\begin{table}[htbp]
\caption{Evaluation results of our AI marker extraction model}
\label{tab:tab3}
\begin{tabular}{lcccc}
\toprule
                  & \multicolumn{2}{c}{Our model} & \multicolumn{2}{c}{CNN+BiLSTM+CRF} \\ \cmidrule(lr){2-3} \cmidrule(lr){4-5}
Type              & Original      & Lowercase     & Original        & Lowercase        \\ \midrule
F1                & 0.792         & 0.772         & 0.784           & 0.737            \\
Recall            & 0.782         & 0.762         & 0.779           & 0.701            \\
Precision         & 0.802         & 0.783         & 0.789           & 0.776            \\
Method-F1         & 0.791         & 0.783         & 0.773           & 0.712            \\
Method-Recall     & 0.804         & 0.780         & 0.761           & 0.683            \\
Method-Precision  & 0.791         & 0.782         & 0.780           & 0.753            \\
Dataset-F1        & 0.802         & 0.773         & 0.794           & 0.761            \\
Dataset-Recall    & 0.813         & 0.753         & 0.784           & 0.742            \\
Dataset-Precision & 0.803         & 0.802         & 0.802           & 0.790            \\
Metric-F1         & 0.770         & 0.741         & 0.740           & 0.782            \\
Metric-Recall     & 0.683         & 0.712         & 0.731           & 0.714            \\
Metric-Precision  & 0.882         & 0.783         & 0.762           & 0.851            \\
\bottomrule
\end{tabular}
\end{table}

\subsubsection{AI markers normalisation}

A series of rule strategies is formulated to normalise the AI markers that have multiple representations. For example, "LSTM", "LSTM-based" and "Long Short-Term Memory" are normalised into "(LSTM)Long Short-Term Memory". All metrics, including "accuracy", such as "mean accuracy" and "predictive accuracy", are normalised into "accuracy". The detailed normalisation strategies are shown in Appendix A. Many AI markers can be distinguished according to the entity types, and the probability of polysemous occurrences in the same type of AI markers is small. Hence, the cases of polysemous entities are not specifically dealt with.

\subsection{Original paper tracing for AI markers}
We must trace back to the original papers of the method or dataset to obtain the research trace of a method or dataset that has been gradually cited by other literature. Only the methods and datasets that clearly appear in the "methodology" or "experiment" chapters of the cited papers are traced.

\subsubsection{Tracing approach}
In a paper, when a method or dataset is cited, the references for the corresponding original papers are often attached to it. In tracing, the set of papers citing the AI marker is firstly recorded for each AI marker. The sentences where the AI marker appears are located for each paper in the set. In each sentence, the existence of references in one or two positions behind the AI marker is checked. If a reference is present, then it is recorded. Finally, the most cited paper corresponding to each AI marker in the recorded references is selected as the original paper.

\subsubsection{Evaluation results}
Using the above tracing approach, 5118 methods proposed in CCF corpus that were explicitly cited more than once and 4105 corresponding original papers are obtained. 1265 datasets proposed in CCF corpus that were explicitly cited more than once and 949 corresponding original papers are obtained.

We randomly select 800 methods from the obtained results, including 200 methods that were explicitly cited 5 times, 200 methods that were explicitly cited 4 times, 200 methods that were explicitly cited 3 times and 200 methods that were explicitly cited 2 times. And we randomly select 400 datasets from the obtained results, including 100 datasets that were explicitly cited 5 times, 100 datasets that were explicitly cited 4 times, 100 datasets that were explicitly cited 3 times and 100 datasets that were explicitly cited 2 times. The original paper results corresponding to the 800 methods and 400 datasets are manually evaluated. The evaluation results are shown in Table~\ref{tab:tab4}. The accuracies exceeding 90\%.

\begin{table}[htbp]
\caption{Evaluation results of the tracing approach}
\label{tab:tab4}
\begin{tabular}{ccccc}
\toprule
                         & Cited & Evaluation & Right & Accuracy \\ \midrule
\multirow{4}{*}{Method}  & 5     & 200        & 196   & 98\%     \\
                         & 4     & 200        & 190   & 95\%     \\
                         & 3     & 200        & 188   & 94\%     \\
                         & 2     & 200        & 185   & 92.5\%   \\
\multirow{4}{*}{Dataset} & 5     & 100        & 95    & 95\%     \\
                         & 4     & 100        & 94    & 94\%     \\
                         & 3     & 100        & 91    & 91\%     \\
                         & 2     & 100        & 90    & 90\%     \\ \bottomrule
\end{tabular}
\end{table}

\subsection{Clustering of methods and research scenes}
A single dataset or a single metric may correspond to multiple different research scenes. For example, the combination of the CMU PIE dataset and the accuracy metric is represented as a face recognition research scene. The combination of the IMDB dataset and the accuracy metric is represented as a movie review emotion classification research scene. Therefore, a dataset and a metric in the same literature is combined to represent a research scene in our work. A larger number of candidate and redundant research scenes can be obtained.

Many metrics are simultaneously applied on the collected CCF corpus, such as precision and recall. Therefore, metrics need to be merged to reduce the redundancy of research scenes.

A method-scene matrix based on the number of co-occurrences of methods and research scenes in the literature is constructed. The number of research scenes is large due to the considerable combinations of datasets and metrics, thereby resulting in the high-dimensional sparseness of the method-scene matrix. To solve this problem, Nonnegative Matrix Factorization (NMF) \cite{r26,r27} and spectral clustering \cite{r28} are used together to build dimensionality reduction and clustering algorithms.

Firstly, the datasets and metrics are combined into research scenes, and then the method-research scene co-occurrence matrix is obtained according to the co-occurrence relationships amongst the methods and research scenes. Secondly, the methods are clustered on the basis of NMF and spectral clustering, and 500 method clusters are obtained. Then, a method cluster-metric co-occurrence matrix is constructed, and only the metrics are clustered by spectral clustering. Fifty metric clusters are obtained. The metric clusters and datasets are combined into research scenes. The research scenes are clustered by spectral clustering according to the method-research scene co-occurrence matrix, and 500 research scene clusters are obtained. We expect that the number of research scenes in each cluster is roughly balanced; hence, the clusters containing more than 500 research scenes will be clustered again by spectral clustering according to the method-research scene co-occurrence matrix. Two clusters have more than 500 research scenes, and 200 research scene clusters are obtained by spectral clustering again. 698 research scene clusters are obtained by combining the 200 research scene clusters with the remaining 498 research scene clusters\footnote{The number of clusters is the experimental parameters. It was verified by manually checking the clusters that the results are reasonable when the number of clusters are the above values.}.

\subsection{Roadmap generation for method clusters}
A method roadmap describes the evolution of different yet highly correlated methods \cite{r29}. In the method clusters obtained by the clustering algorithm, each method cluster is composed of the same type of methods. In the cluster, if a time-based method evolution map can be built, and the dataset information can be added, then it will provide enlightening information for related research.

The roadmap generation procedure for a method cluster (the output of the algorithm described in Section 3.4) proposed in the study is as follows:

1) The information of all the methods in the cluster is obtained, including the time of proposal, the chapters where the method locates in its original paper and the associated datasets used in its original paper\footnote{If the original paper is not traced back in Section 3.3, then the earliest paper cited by this method will be used as the original paper.};

2) For each method ${M_i}$  in the cluster, the other methods $\left\{ {{M_1},...,{M_n}} \right\}$ mentioned in the "experiment" chapter of the original paper of ${M_i}$  are found. Path ${M_i} \to {M_j}$ from ${M_i}$  to ${M_j}$  is constructed, and ${M_j} \in \left\{ {{M_1},...,{M_n}} \right\}$. The edge between ${M_i}$ and ${M_j}$ represents the datasets used for the comparison between ${M_i}$ and ${M_j}$;

3) Continuous paths are combined to obtain the roadmap of the methods in the same method cluster. For example, if $\left( {{M_1} \to {M_2}} \right)$, $\left( {{M_2} \to {M_3}} \right)$ and $\left( {{M_1} \to {M_3}} \right)$ exist, then only $\left( {{M_1} \to {M_2}} \right)$ and $\left( {{M_2} \to {M_3}} \right)$ are kept.

The construction of our roadmaps differs from the approach in \cite{r29} in two points: 1) Datasets are added in our roadmap construction, which provides additional information; 2) The methods are obtained by large-scale literature mining in our roadmap, and numerous roadmaps can be simultaneously obtained.

\subsection{Influencing degrees for research scene clusters}
The influencing degrees amongst the research scene clusters in the literature and that of the effective methods proposed by the literature on other research scene clusters are analysed.

The research scene cluster corresponding to the research scene involved in each paper is found. Considering that a paper generally only involves one major research scene, the research scene cluster with the most occurrences of each paper is taken as the research scene cluster corresponding to the paper. In the end, the research scene clusters corresponding to 45,215 papers in CCF corpus are obtained\footnote{Given that the remaining papers in the CCF corpus are not extracted datasets or metrics, they cannot form research scenes, and no corresponding research scene clusters are present.}. The effective methods proposed by 45,215 papers and the corresponding original papers are combined. Thereafter, the influencing degrees amongst the research scene clusters in these 45,215 papers and that of the effective methods proposed by the 45,215 papers on other research scene clusters are analysed.

Variable ${L_s}$ is the set of papers whose research scene cluster is $s$, $l_s^i \in {L_s}$; and ${L_{\backslash s}}(i)$ is the set of papers whose scene cluster is not $s$  and that proposes effective methods that are cited by $l_s^i$  within three years, $l_{\backslash s}^j(i) \in {L_{\backslash s}}(i)$. The influencing degree rate of scene cluster $\backslash s$ to scene cluster $s$ is calculated as follows:

\begin{equation}\label{1}
ID{R_s} = \frac{{\sum\limits_{{\rm{i = 1}}}^{\left| {{L_s}} \right|} {\sum\limits_{j = 1}^{\left| {{L_{\backslash s}}(i)} \right|} {JS\left( {d\left( {l_s^i} \right),d\left( {l_{\backslash s}^j\left( i \right)} \right)} \right)} } }}{{\sum\limits_{i = 1}^{\left| {{L_s}} \right|} {\left| {{L_{\backslash s}}(i)} \right|} }},
\end{equation}
where $d\left( {l_s^i} \right)$ is the distribution of the research scene cluster of the paper $l_s^i$  in 45,215 papers, $d\left( {l_{\backslash s}^j\left( i \right)} \right)$  is the distribution of the research scene cluster of the paper $l_{\backslash s}^j(i)$  in 45,215 papers, and $JS\left( {d\left( {l_s^i} \right),d\left( {l_{\backslash s}^j\left( i \right)} \right)} \right)$  is the Jensen-Shannon divergence of $d\left( {l_s^i} \right)$ and $d\left( {l_{\backslash s}^j\left( i \right)} \right)$.

The influence of the effective methods proposed by the 45,215 papers to other research scene clusters is analysed.

Variable ${l_m}$ is the original paper corresponding to effective method $m$; $s$ is the research scene cluster corresponding to literature ${l_m}$; ${L_{\backslash s}}(m)$ is the set of papers whose scene cluster is not $s$  and cites effective method $m$ within three years, and $l_{\backslash s}^k(m) \in {L_{\backslash s}}(m)$. The influencing degree of effective method $m$ to other research scene clusters is calculated using Formula (2). Meanwhile, the influencing degree rate of effective method $m$ to other research scene clusters is calculated using Formula (3).
\begin{equation}\label{2}
I{D_m} = \sum\limits_{{\rm{k = 1}}}^{\left| {{L_{\backslash s}}(m)} \right|} {JS\left( {d\left( {{l_m}} \right),d\left( {l_{\backslash s}^k\left( m \right)} \right)} \right)} ,
\end{equation}

\begin{equation}\label{3}
ID{R_m} = \frac{{I{D_m}}}{{\left| {{L_{_{\backslash s}}}(m)} \right|}},
\end{equation}
where $d({l_m})$ is the distribution of the research scene cluster of the paper ${l_m}$ in 45,215 papers, $d\left( {l_{\backslash s}^k\left( m \right)} \right)$ is the distribution of the research scene cluster of the paper $l_{\backslash s}^k\left( m \right)$ in 45,215 papers, and  $JS\left( {d\left( {{l_m}} \right),d\left( {l_{\backslash s}^k\left( m \right)} \right)} \right)$ is the Jensen-Shannon divergence of $d({l_m})$ and $d\left( {l_{\backslash s}^k\left( m \right)} \right)$.

\section{Results}
This section performs statistical analysis, propagation analysis and mining on the basis of the AI markers in the collected CCF corpus (2005-2019 AI papers). The results are displayed on the basis of the aforementioned approaches, including chapter classification, AI marker extraction and normalisation, original paper tracing for AI markers, clustering of methods and research scenes, roadmap generation for the method clusters and influencing degrees for research scene clusters.

\subsection{Statistics of effective AI markers}
171,677 machine learning method entities, 16,645 dataset entities and 1551 metric entities are mined by extracting the AI markers in the CCF corpus. Only AI markers that are cited more than once are considered in the analysis. The AI markers that appear more than once are called effective AI markers.

This section introduces the analysis of effective AI markers in terms of the countries and publication venues. The top 10 effective AI markers used every year are also described.

\subsubsection{Analysis in terms of countries}
\begin{figure}[htbp]
\setlength{\belowcaptionskip}{-0.3cm}
    \centering
    \includegraphics[width=\linewidth]{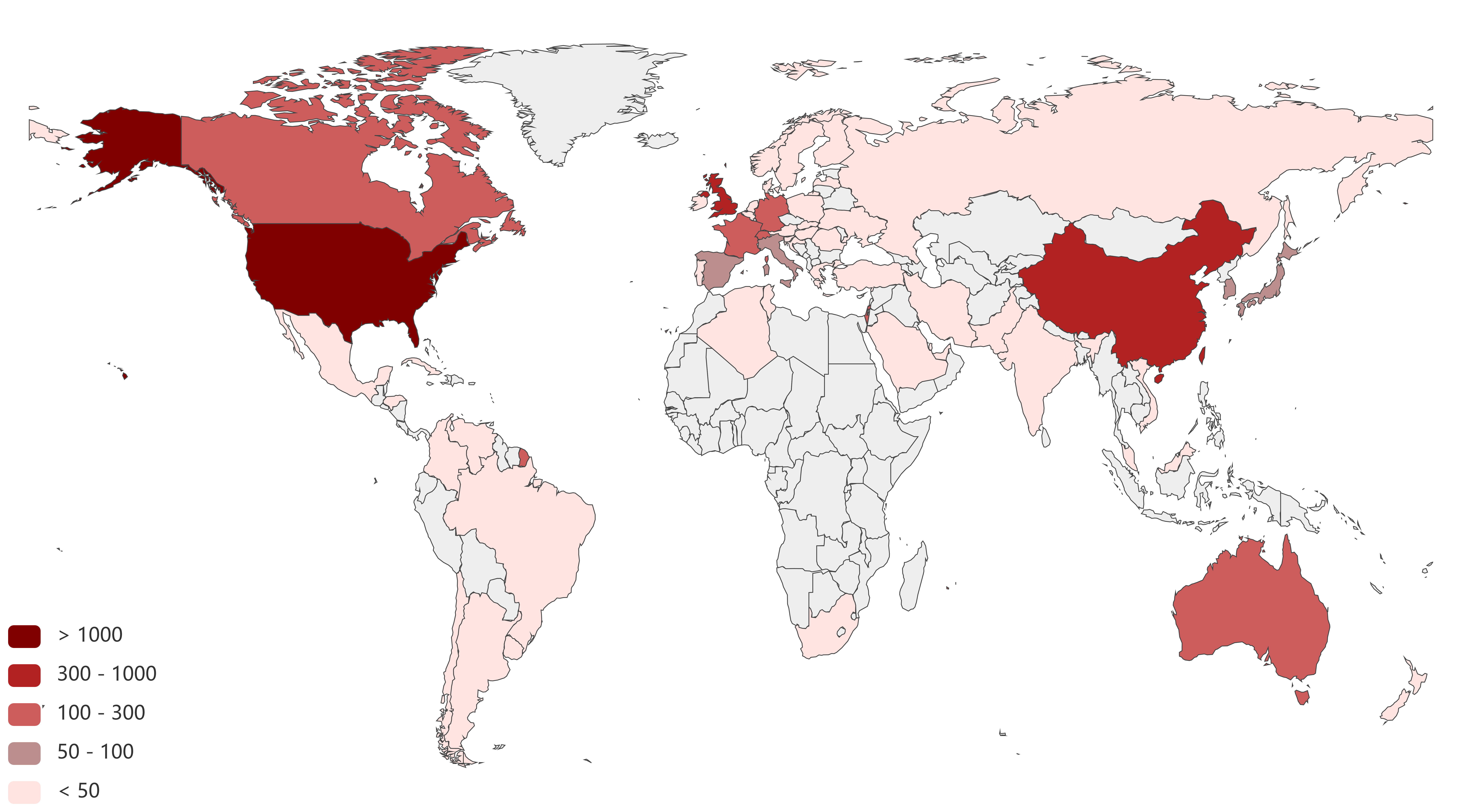}
    \caption{Distribution of the number of effective methods in different countries.}
    \label{fig:fig4}
\end{figure}

\begin{figure}[htbp]
\setlength{\belowcaptionskip}{-0.3cm}
    \centering
    \includegraphics[width=\linewidth]{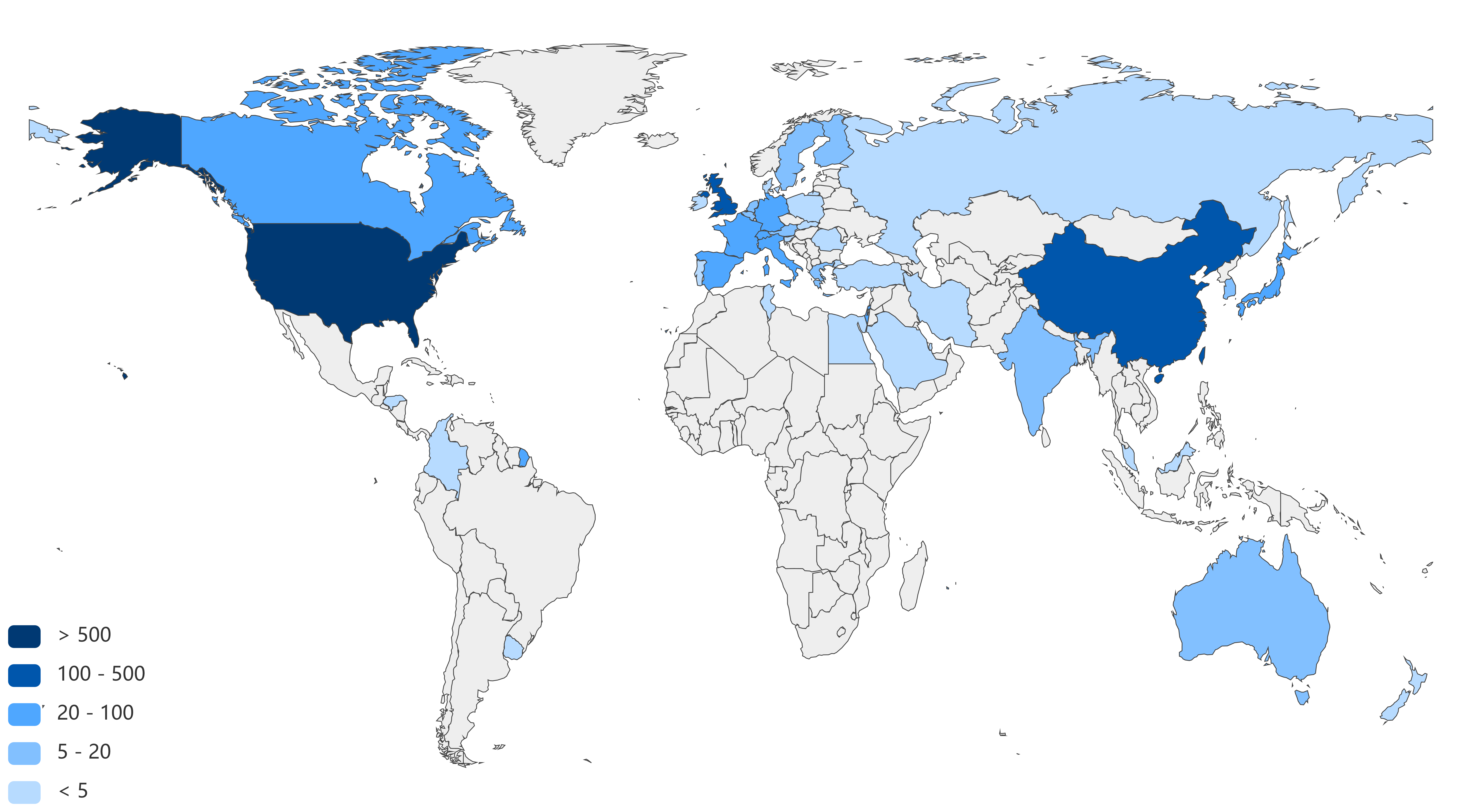}
    \caption{Distribution of the number of effective datasets in different countries.}
    \label{fig:fig5}
\end{figure}

The number of effective AI markers proposed by a country can partially reflect its AI research level . The number of effective methods and datasets proposed by each country in the CCF corpus from 2005 to 2019 are calculated (Fig. \ref{fig:fig4} and \ref{fig:fig5}).

Figure \ref{fig:fig4} demonstrates that the top three countries according to effective method quantities are the United States, China and the United Kingdom, followed by Germany, France, Canada, Singapore, Australia and so on. In Fig. \ref{fig:fig5}, the top three countries according to effective dataset quantities are also the United States, China and the United Kingdom, followed by Germany, Switzerland, Canada, France, Singapore, Israel and so on. The United States, China and the United Kingdom are relatively active countries in the field of AI. Although countries, such as Germany, France, Canada and Singapore, have a certain gap with the United States, China and the United Kingdom, they are also relatively active.

The proposal rates of the effective methods and datasets for each country are calculated to reduce the effect of the number of papers published in each country.

The proposal rate $M{R_c}$ of the effective methods of country $c$ and the proposal rate  $D{R_c}$ of effective the datasets of country $c$ are calculated using Formulas (4) and (5).

\begin{equation}\label{4}
M{R_c} = \frac{{\left| {{M_c}} \right|}}{{\left| {{L_c}} \right|}},
\end{equation}

\begin{equation}\label{5}
D{R_c} = \frac{{\left| {{D_c}} \right|}}{{\left| {{L_c}} \right|}},
\end{equation}
where ${M_c}$ is the set of effective methods proposed by country $c$, ${D_c}$ is the set of effective datasets proposed by country $c$ , and ${L_c}$ is the set of papers proposed by country $c$.

\begin{figure}[htbp]
\setlength{\belowcaptionskip}{-0.3cm}
\centering
\subfigure[Proposal rates of the effective methods of the top 10 countries listed in Fig. \ref{fig:fig4}.]{
\begin{minipage}[b]{0.6\textwidth}
\includegraphics[width=1\textwidth]{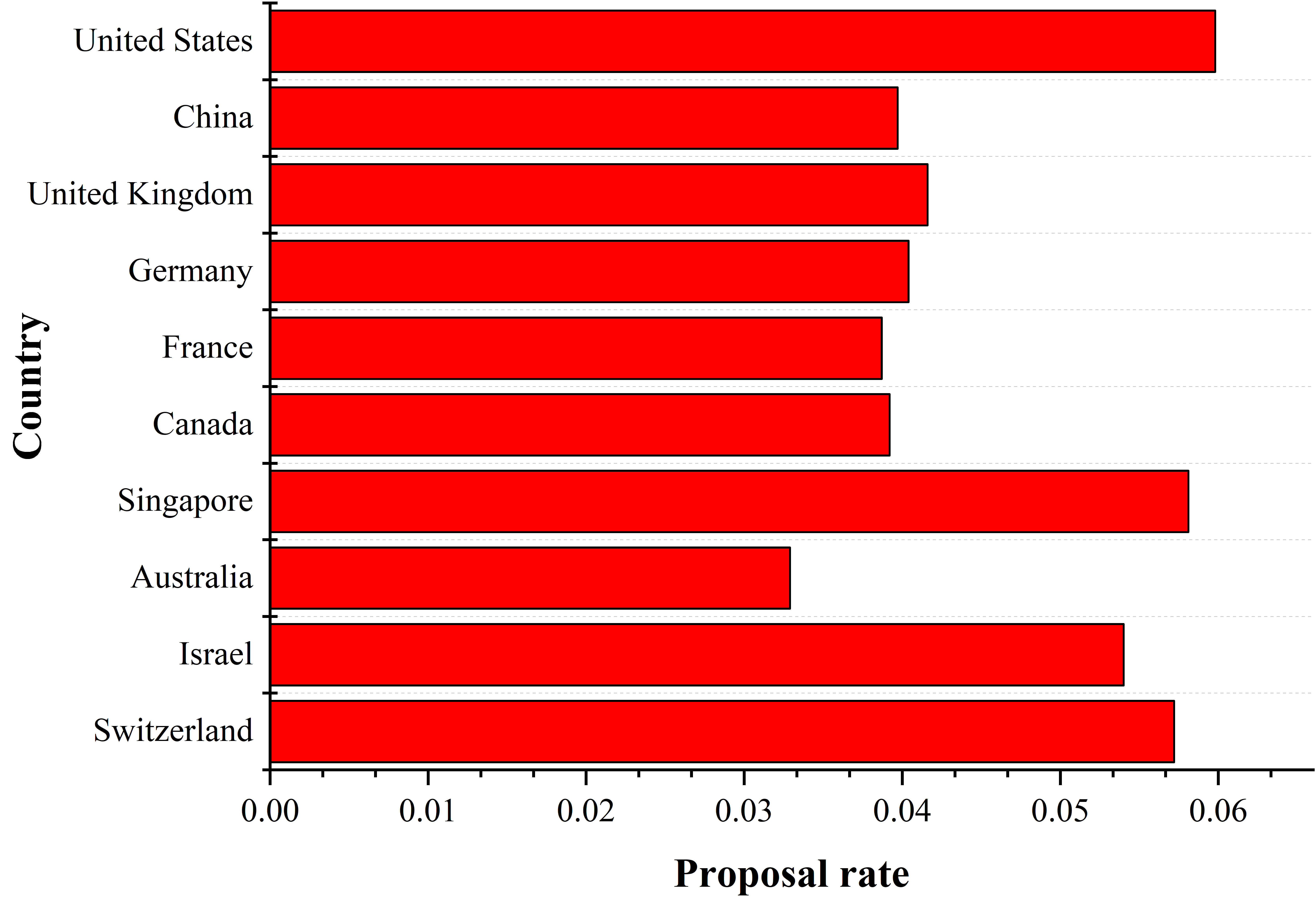}
\label{fig:fig6a}
\end{minipage}
}

\subfigure[Proposal rates of the effective datasets of the top 10 countries listed in Fig. \ref{fig:fig5}.]{
\begin{minipage}[b]{0.6\textwidth}
\includegraphics[width=1\textwidth]{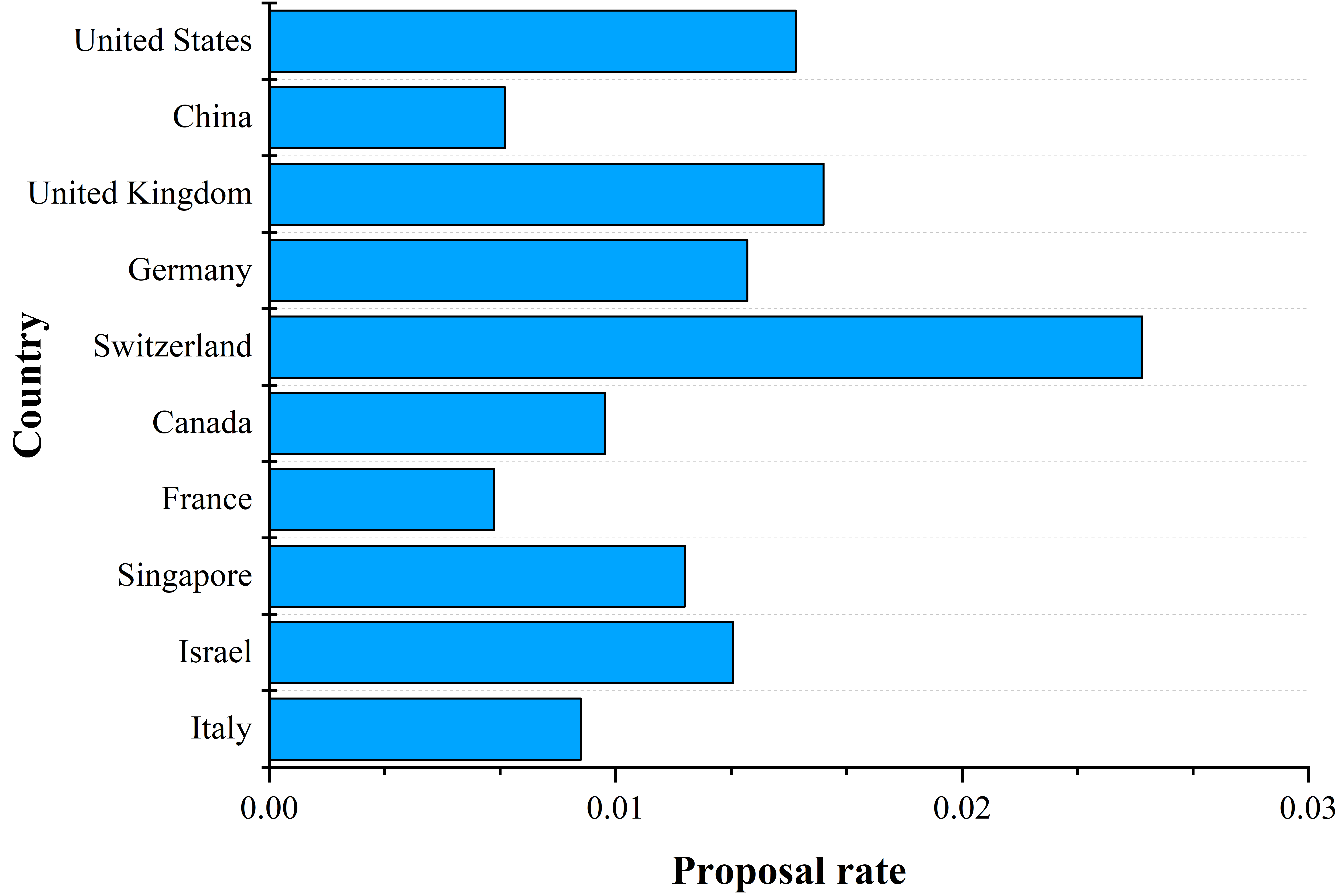}
\label{fig:fig6b}
\end{minipage}
}

\caption{Proposal rates of the effective AI markers of the top 10 countries listed in Fig. \ref{fig:fig4} and Fig. \ref{fig:fig5}. The number of effective AI markers proposed by the countries decreased from top to bottom.} \label{fig:fig6}
\end{figure}

The proposal rates of the effective methods of the top 10 countries in terms of the number of proposed effective methods and those of the effective datasets of the top 10 countries with regard to the number of proposed effective datasets are obtained on the basis of Formulas (4) and (5). The results are shown in Fig. \ref{fig:fig6}.

In Fig. \ref{fig:fig6a}, the proposal rate of the effective methods of the United States ranked first. The rates of China and the United Kingdom are lower than those of Singapore, Israel and Switzerland. In Fig. \ref{fig:fig6b}, the proposal rate of Switzerland is the highest even though its number of proposed effective datasets is lower than that of the United States, China, the United Kingdom and Germany. This result reflects that Switzerland attaches great importance to AI datasets.

\subsubsection{Analysis in terms of the publication venues}
The number of effective AI markers proposed by a publication venue can reflect the quality of the publication venue. The proposal rate $M{R_v}$ of the effective methods of publication venue $v$ and the proposal rate $D{R_v}$ of the effective datasets of publication venue $v$ are calculated according to Formulas (6) and (7).
\begin{equation}\label{6}
M{R_v} = \frac{{\left| {{M_v}} \right|}}{{\left| {{L_v}} \right|}},
\end{equation}

\begin{equation}\label{7}
D{R_v} = \frac{{\left| {{D_v}} \right|}}{{\left| {{L_v}} \right|}},
\end{equation}
where ${M_v}$ is the set of effective methods proposed by publication venue $v$, ${D_v}$ is the set of effective datasets proposed by publication venue $v$ , and ${L_v}$ is the set of papers proposed by publication venue $v$.

\begin{figure}[htbp]
\setlength{\belowcaptionskip}{-0.3cm}
\centering
\subfigure[Proposal rates of the effective methods of the top 10 publication venues in terms of the number of proposed effective methods.]{
\begin{minipage}[b]{0.6\textwidth}
\includegraphics[width=1\textwidth]{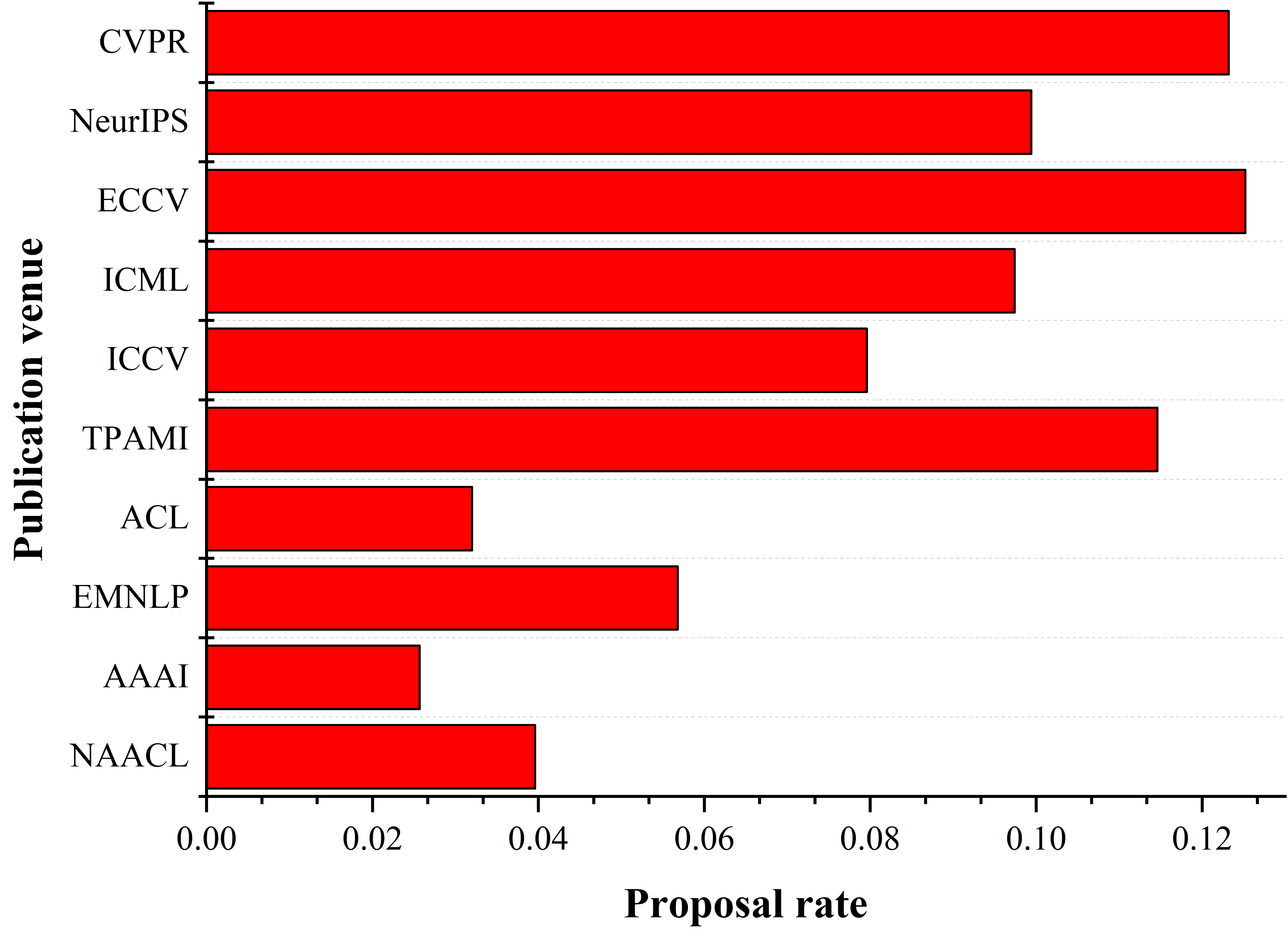}
\label{fig:fig7a}
\end{minipage}
}

\subfigure[Proposal rates of the effective datasets of the top 10 publication venues in terms of the number of proposed effective datasets.]{
\begin{minipage}[b]{0.6\textwidth}
\includegraphics[width=1\textwidth]{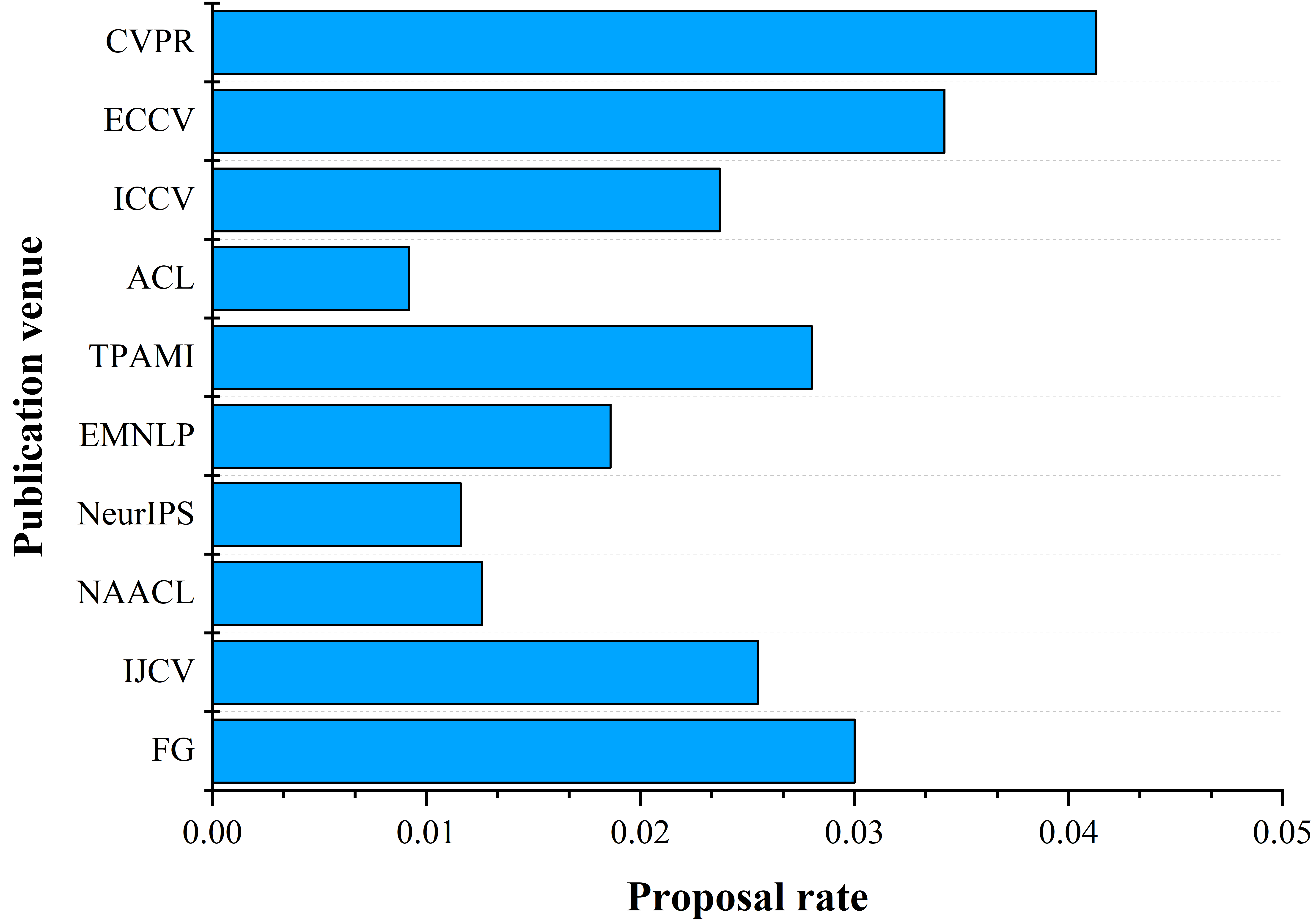}
\label{fig:fig7b}
\end{minipage}
}

\caption{Proposal rates of the effective AI markers of the top 10 publication venues in terms of the number of proposed effective AI markers. The number of effective AI markers proposed by publication venues decreased from top to bottom.} \label{fig:fig7}
\end{figure}

The proposal rates of the effective methods of the top 10 publication venues in terms of the number of proposed effective methods and those of the effective datasets of the top 10 publication venues with regard to the number of proposed effective datasets are obtained on the basis of on Formulas (6) and (7). The results are shown in Fig. \ref{fig:fig7}.

In Fig. \ref{fig:fig7a}, the proposal rate of the effective methods of ECCV is higher than that of CVPR even though it ranked B in the CCF. Amongst the top 10 publication venues in terms of the number of proposed effective methods, seven publication venues belong to Tier-A in CCF. The quality of papers in the publication venues of Tier-A is indeed higher than that of Tier-B and Tier-C.

Figure \ref{fig:fig7b} shows the distribution of the effective datasets. The proposal rate of the effective datasets of CVPR ranked first. Although ECCV belongs to Tier-B in CCF, the proposal rate of the effective datasets is second only to CVPR. Six publication venues belong to Tier-A in the top 10 publication venues in terms of the number of proposed effective datasets. This finding reflects that Tier-A publication venues pay great attention to the proposal of effective datasets.

\subsubsection{Annual top 10 effective AI markers}
The section analyses the number of effective methods and datasets used every year from 2005 to 2019.

\textbf{(1) Top 10 effective methods}

\begin{figure}[htbp]
\setlength{\belowcaptionskip}{-0.3cm}
    \centering
    \includegraphics[width=\linewidth]{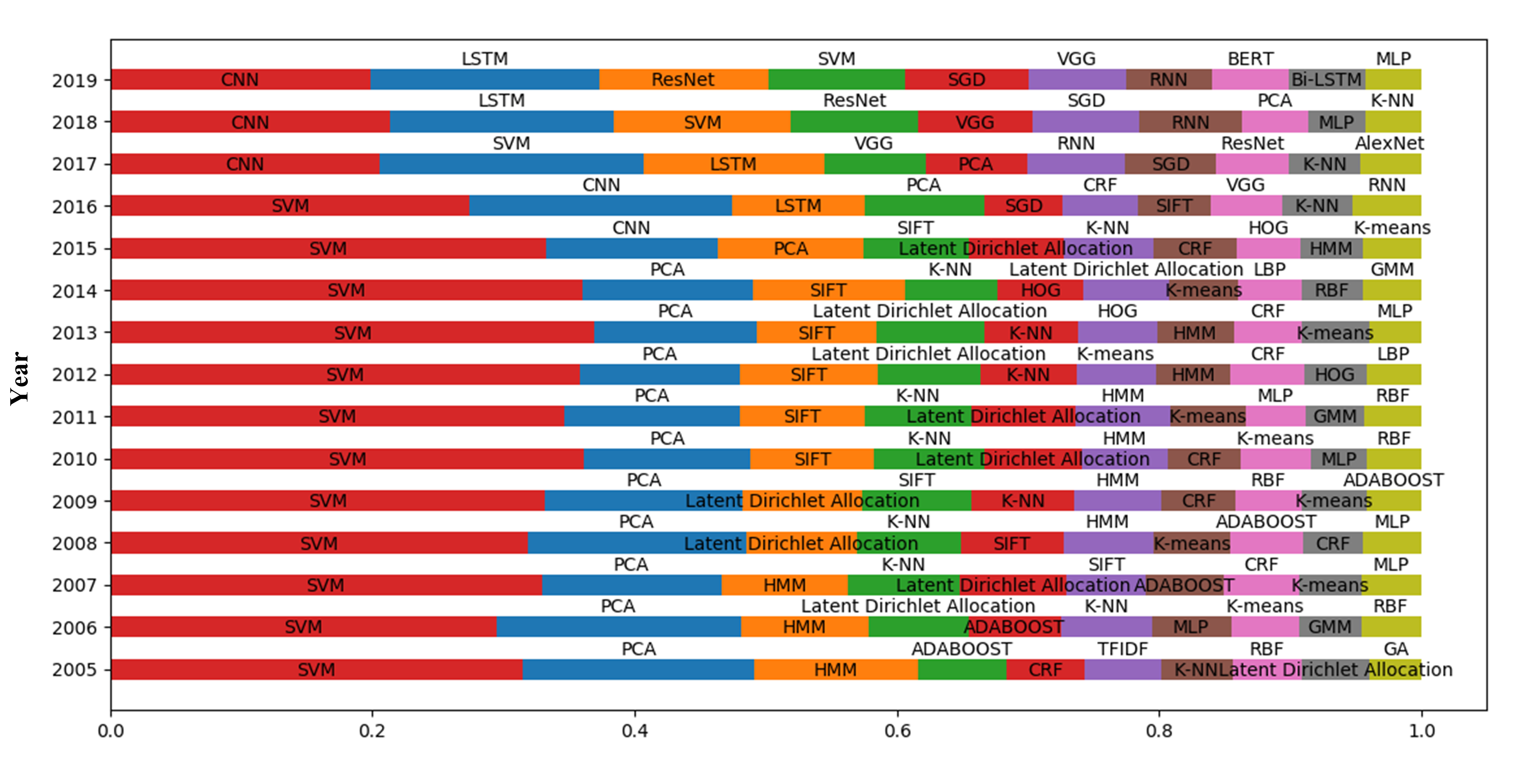}
    \caption{Top 10 effective methods used every year.}
    \label{fig:fig8}
\end{figure}

The number of effective methods used every year from 2005 to 2019 is counted. The top 10 popular effective methods used every year are shown in Fig. \ref{fig:fig8}.

In Fig. \ref{fig:fig8}, SVM, a traditional machine learning method, is widely used every year. LDA, a classical topic model for text mining, was widely used from 2005 to 2015. However, the proportion of LDA usage significantly decreased after 2015 with the rapid development of deep learning. After 2015, deep learning becomes increasingly popular, and deep learning methods have become the mainstream in AI.

Computer vision and natural language processing are two important research disciplines in AI research. Fig. \ref{fig:fig8} demonstrates that the methods in the computer vision have always occupied a large proportion, thus indicating that computer vision has always been a popular research branch of AI.

\textbf{(2) Top 10 effective datasets used every year}

\begin{figure}[htbp]
\setlength{\belowcaptionskip}{-0.3cm}
    \centering
    \includegraphics[width=\linewidth]{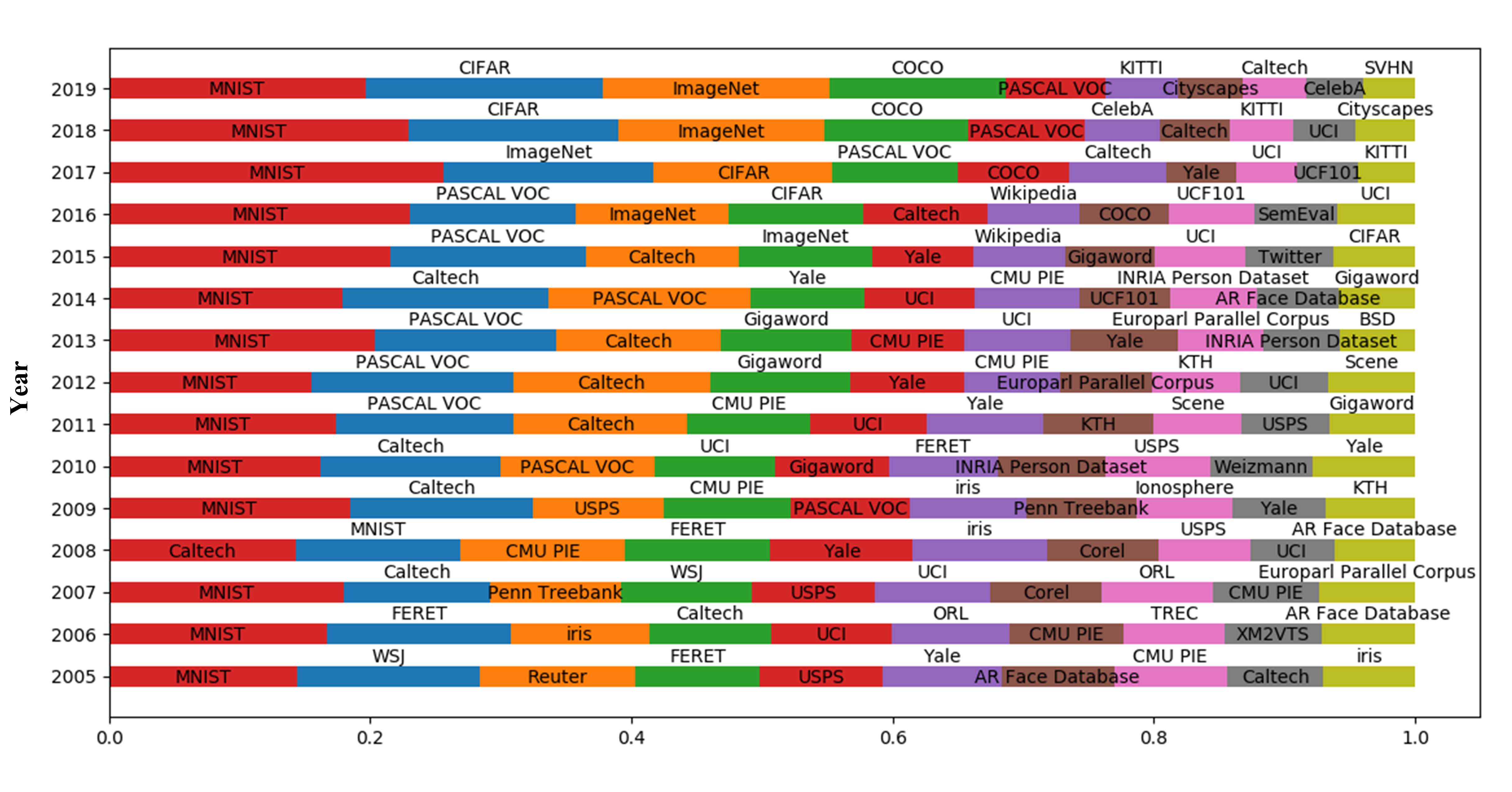}
    \caption{Top 10 effective datasets used every year.}
    \label{fig:fig11}
\end{figure}

The number of effective datasets used every year is counted. The top 10 effective datasets used every year are shown in Fig. \ref{fig:fig11}.

MNIST is commonly used every year. In 2016, SemEval dataset entered the top 10 rank. This dataset is mainly used for sentiment analysis. In 2016, sentiment analysis became a major research field. In 2017, the KITTI dataset entered the top 10 rank. This dataset is mainly used in the field of driverless, thus indicating that driverless became a major research field in 2017. The proportion of the KITTI dataset in the top 10 datasets gradually increased from 2017 to 2019. The datasets will generally take at least 2 years to be recognised and widely used in the corresponding field since it was proposed. For example, PASCAL VOC dataset was proposed in 2007 and widely used in 2009. Weizmann dataset was proposed in 2006 and widely used in 2010. COCO dataset was proposed in 2014 and widely used in 2016.

Face recognition is a popular research direction in the computer vision field. The effective method proportion of face recognition in the top 10 effective methods used every year is counted (Table~\ref{tab:tab5}).

\begin{table}[htbp]
\caption{Proportion of effective datasets of face recognition in the top 10 effective datasets used every year}
\label{tab:tab5}
\begin{tabular}{ccc}
\toprule
Year & Face recognition                                         & Proportion \\ \midrule
2005 & FERET, Yale, AR Face Database, CMU PIE and Caltech        & 43.2\%     \\
2006 & FERET, Caltech, ORL, CMU PIE, XM2VTS and AR Face Database & 55.7\%     \\
2007 & Caltech, ORL and CMU PIE                                    & 28.0\%     \\
2008 & Caltech, CMU PIE, FERET, Yale and AR Face Database        & 55.1\%     \\
2009 & Caltech, CMU PIE and Yale                                   & 30.9\%     \\
2010 & Caltech, FERET and Yale                                     & 30\%       \\
2011 & Caltech, CMU PIE and Yale                                   & 31.6\%     \\
2012 & Caltech, Yale and CMU PIE                                   & 31.2\%     \\
2013 & Caltech, CMU PIE and Yale                                   & 29.4\%     \\
2014 & Caltech, Yale, CMU PIE and AR Face Database                 & 38.7\%     \\
2015 & Caltech and Yale                                            & 19.4\%     \\
2016 & Caltech                                                  & 9.5\%      \\
2017 & Caltech and Yale                                            & 12.8\%     \\
2018 & CelebA and Caltech                                          & 11.1\%     \\
2019 & Caltech and CelebA                                          & 9.2\%      \\ \bottomrule
\end{tabular}
\end{table}

The datasets commonly used for face recognition from 2005 to 2019 are Caltech, Yale, CMU PIE and CelebA. Caltech appears in the top 10 effective datasets every year and has a high proportion. Yale has appeared in many years; however, after the CelebA dataset appeared, its status was replaced by CelebA.

\subsection{Propagation of effective methods}
The section analyses the propagation on the datasets and that amongst countries with effective methods\footnote{Given that the propagation analysis needs to be traced to the original papers of the effective methods, the effective methods and the corresponding original papers obtained in Section 3.3 are only used for analysis in this section.}.
\subsubsection{Propagation on datasets}
The propagation rates of the effective methods proposed in year $y$ on the datasets in the time period from year  $y$ to year $y + \Delta y$ is calculated using Formula (8).

\begin{equation}\label{8}
PR\left( {\Delta y|y} \right) = \frac{{\sum\limits_{m \in {M_y}} {\left| {{D_m}\left( {\Delta y|y} \right)} \right|} }}{{\left| {{M_y}} \right|}},
\end{equation}
where ${M_y}$ is the set of all the methods proposed in year $y$,  ${D_m}\left( {\Delta y|y} \right)$ is the set of the datasets applied to method $m$ in the time period from year $y$ to year $y + \Delta y$ , and $\Delta y \in \{ 0,1,2\}$.

\begin{figure}[htbp]
\setlength{\belowcaptionskip}{-0.3cm}
    \centering
    \includegraphics[scale=0.4]{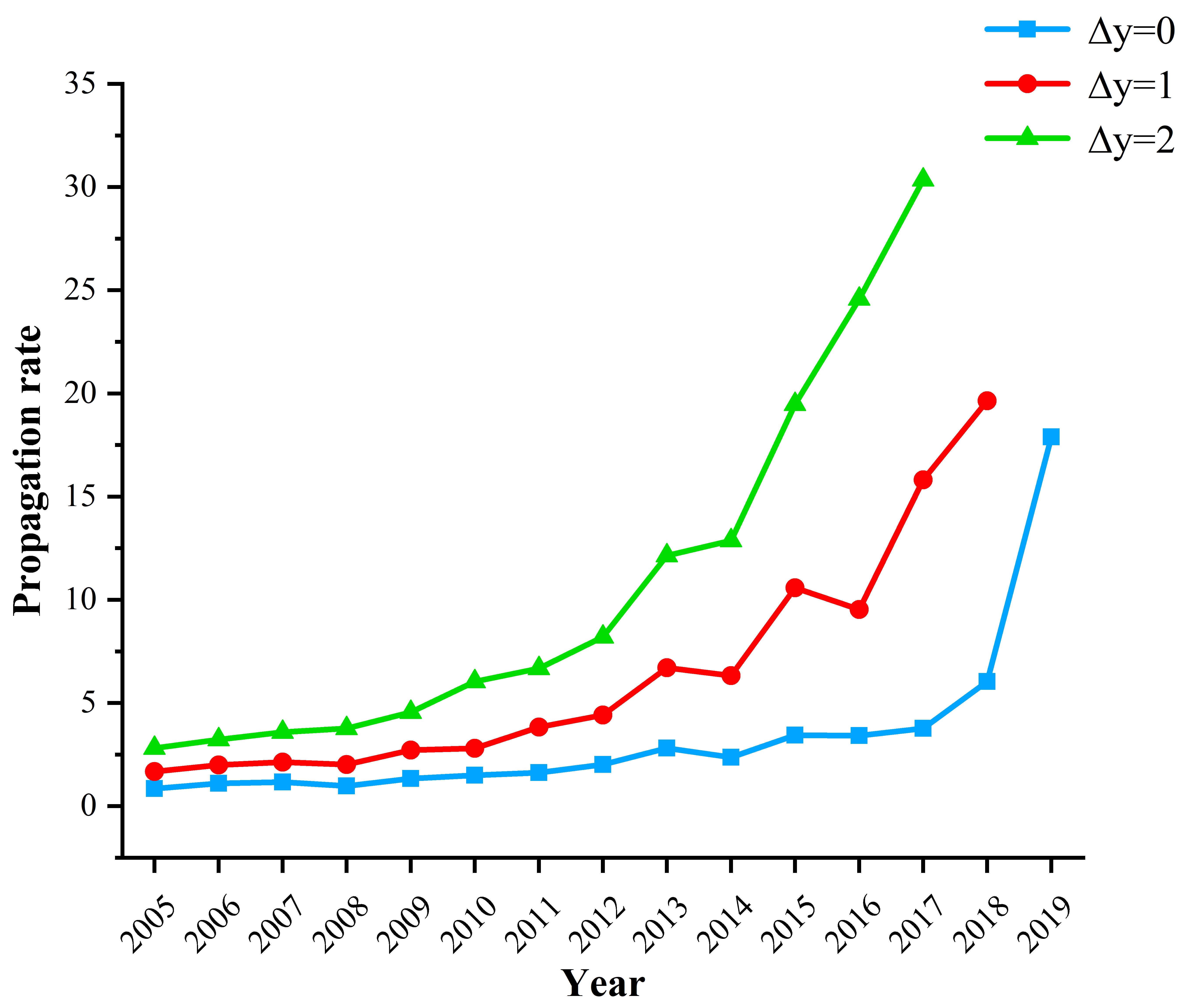}
    \caption{Propagation rates of the effective methods on the datasets.}
    \label{fig:fig12}
\end{figure}

Based on Formula (8), the propagation rates of the effective methods on the datasets are obtained and shown in Fig. \ref{fig:fig12}. The aforementioned figure shows that the propagation rate gradually increases with the development of time. Various well-known methods can be accessed through various channels, such as arxiv.

\begin{figure}[htbp]
\setlength{\belowcaptionskip}{-0.3cm}
\centering

\subfigure[Application of LMNN to the datasets in 2006 (inner circle) and 2007]{

\begin{minipage}[b]{0.5\textwidth}

\includegraphics[width=1\textwidth]{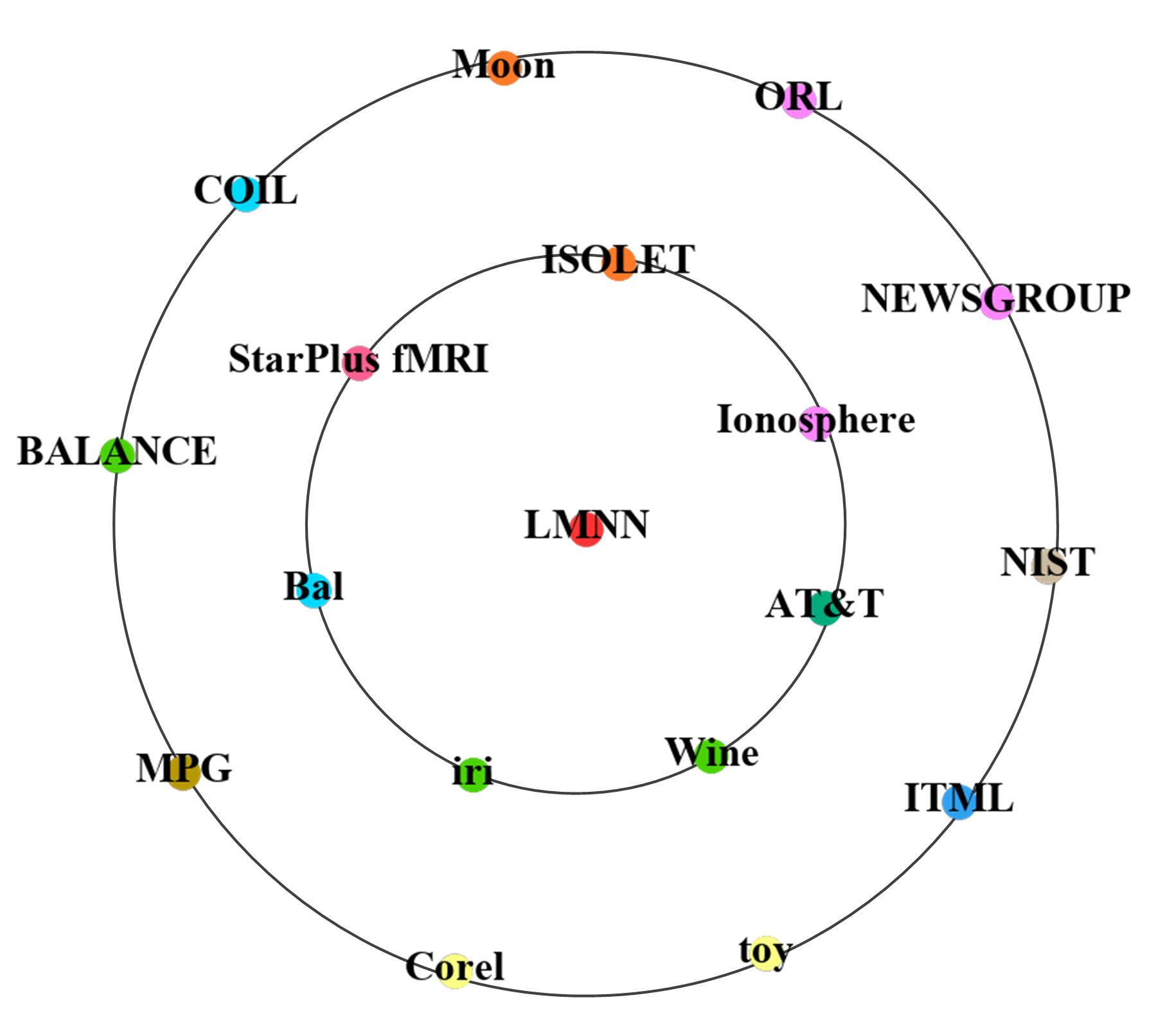}
\label{fig:fig13a}
\end{minipage}

}

\subfigure[Application of Transformer to the datasets in 2018 (inner circle) and 2019]{

\begin{minipage}[b]{0.5\textwidth}

\includegraphics[width=1\textwidth]{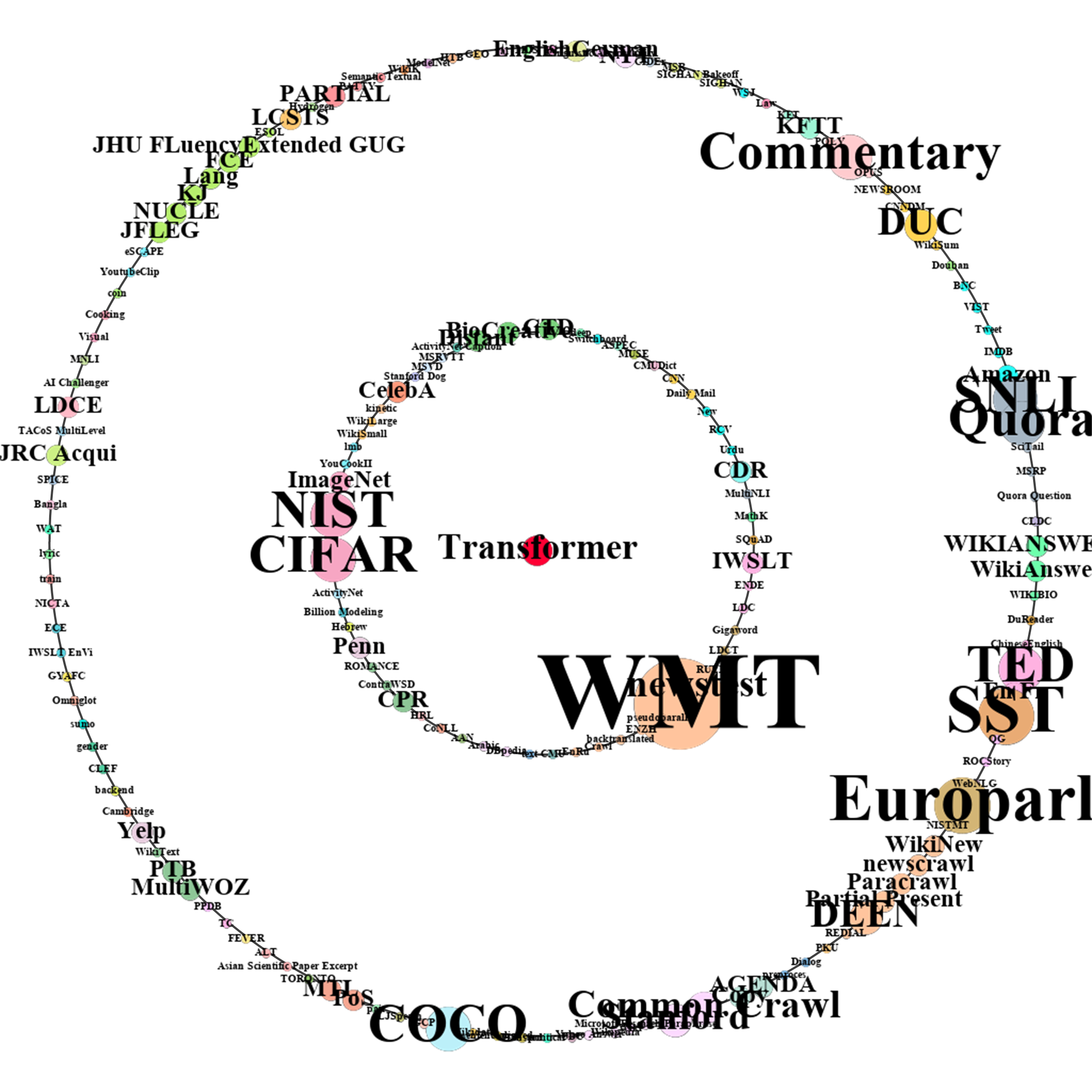}
\label{fig:fig13b}
\end{minipage}

}

\caption{Propagation rates of the effective methods on the datasets. The red point in the centre indicates the method. The inner and outer circles are composed of many dataset points. In the dataset points, the size of the point indicates the number of the dataset applied to the method. The colours of the different dataset points indicate various research scenes.} \label{fig:fig13}

\end{figure}

We also compare the application of different methods on the datasets within two years after being propagated to other literature. The Large margin nearest neighbour (LMNN) and Transformer methods proposed in 2005 and 2018, respectively, are considered in the case study. The results are shown in Fig. \ref{fig:fig13}.

Figure \ref{fig:fig13} demonstrates that after Transformer was proposed in 2018, it was applied to many different datasets in 2018 and 2019. However, LMNN, which was proposed in 2005, has been cited in other literature in 2006 and then applied to different datasets. Moreover, the number and types of applications of Transformer on the datasets in two years are far more than those of LMNN. This notion reflects that the propagation of methods on the datasets is rapidly increasing with the development of time.

\subsubsection{Propagation amongst countries}
Variable ${M_c}$  is the set of all the methods proposed by country $c$, and  $ m\in {M_c}$. The propagation degree of the effective methods from country $c$ to country $c'$ in the time period from year $y$ to year  $y + \Delta y$  is calculated using Formula (9).

\begin{equation}\label{9}
P{D_{c,c'}}(\Delta y|y) = \sum\limits_{m \in {M_c}} {1 * \left| {LE_{c'}^m(\Delta y|y)} \right|{\rm{ + }}0.5 * \left| {LM_{c'}^m(\Delta y|y)} \right|},
\end{equation}
where $LE_{c'}^m(\Delta y|y)$ is the set of papers in country $c'$, which cites method m in the "experiment" chapter in the time period from year $y$ to year $y + \Delta y$, $LM_{c'}^m(\Delta y|y)$ is the set of papers in country $c'$, which cites method $m$ in the "methodology" chapter in the time period from year $y$ to year  $y + \Delta y$, $y \in {\rm{\{ }}2005,2006 \cdots 2019{\rm{\} }}$
 and $\Delta y \in {\rm{\{ 0,1,2}} \cdots {\rm{14\} }}$.

\begin{figure}[htbp]
\setlength{\belowcaptionskip}{-0.3cm}
    \centering
    \includegraphics[width=\linewidth]{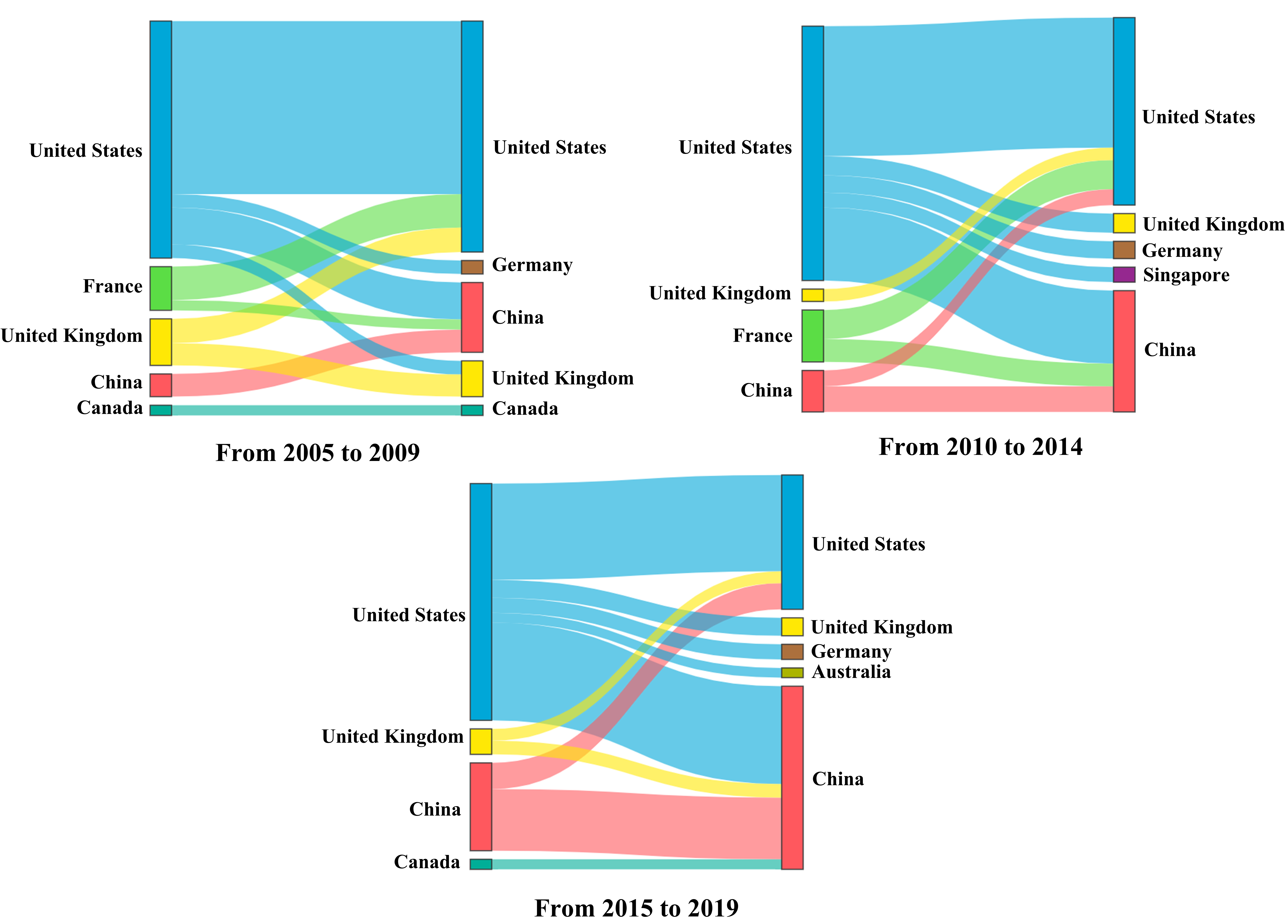}
    \caption{Top 10 propagation degrees of effective methods amongst countries from 2005 to 2019.}
    \label{fig:fig15}
\end{figure}

Based on Formula (9), the propagation degrees of the effective methods amongst countries from 2005 to 2009, from 2010 to 2014 and from 2015 to 2019 are calculated. The top 10 propagation degrees amongst countries at each stage are shown in Fig. \ref{fig:fig15}.

The methods propagated more from the United States, France and the United Kingdom to other countries from 2005 to 2009. The methods proposed by China had a low degree of propagation. From 2010 to 2014, the propagation degree of the methods proposed by China gradually increased. From 2015 to 2019, the propagation degree of the methods proposed by China to the United States ranked fourth place, which reflects that the AI development of China is gradually improving. Meanwhile, the methods proposed by France have a high propagation degree from 2005 to 2014. From 2015 to 2019, the propagation degree of the methods proposed by France ranked after ten, which reflects the relatively slow AI development in France in recent years.

\subsection{Results of roadmap and research scenes}
\subsubsection{Case study for method roadmap}
The following common categories of methods are analysed: representation learning in the knowledge graph and generative adversarial networks. The roadmaps of the methods in the "Trans" and "GAN" clusters are drawn by using our roadmap generation algorithm.

\begin{figure}[htbp]
\setlength{\belowcaptionskip}{-0.3cm}
    \centering
    \includegraphics[width=\linewidth]{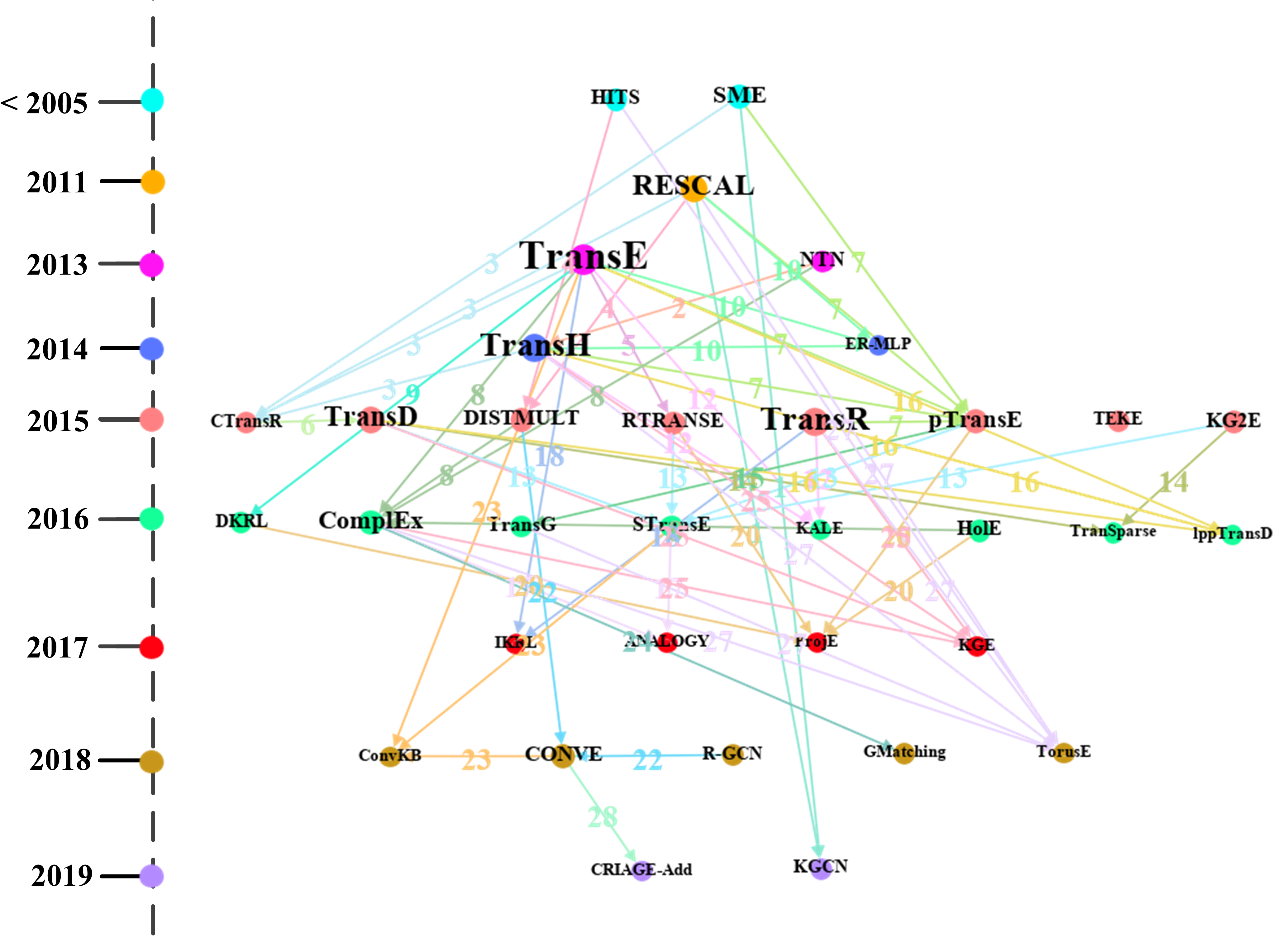}
    \caption{Roadmap of the methods in the "Trans" cluster. The numbers in the figure indicate the datasets used when comparing ${M_i}$ and ${M_j}$ in path ${M_i} \to {M_j}$. The coloured dots in the figure indicate the years, their sizes denote the sizes of the out-degrees, and the coloured lines signify the datasets represented by the numbers.\\The correspondence between the numbers and the datasets is as follows: 1: WIKILINKS; 2: WIKILINKS, WN and FB; 3: WordNet, FB, WN and Freebase; 4: ClueWeb; 5: Family; 6: FB and WN; 7: Freebase, NYT and YORK; 8: WordNet, Freebase and WN; 9:null; 10: RESCAL, WordNet and WN; 11: Freebase; 12: WordNet and Freebase; 13: ClueWeb and WN; 14: FB and WN; 15: WordNet, FB, WN and Freebase; 16: FB and WN; 17: null; 18: KG, ImageNet and WN; 19: null; 20: DBpedia; 21: FB and WN; 22: WN, YAGO and WNRR; 23: WNRR, HIT and MR; 24: Wikione, NELLone and NELL; 25: WNRR and WN; 26: WordNet and WN; 27: WordNet, Freebase and WN; 28: YAGO}
\label{fig:fig16}
\end{figure}

Figure \ref{fig:fig16}  is the generated roadmap of the methods in the "Trans" cluster. After the survey study conducted by Ji et al. \cite{r30} is checked, the roadmap generated from our "Trans" cluster covers 76\% of the representative methods in the knowledge graph mentioned in \cite{r30}, and contains some other representative methods. For example, GMatching and KGE are graph embedding methods, and HITS is a link analytical method. The out-degree of the TransE method is the largest. On the one hand, many methods, such as CTransR and RTRANSE, are inspired by the TransE method and expand new methods. On the other hand, TransE is a representative method of representation learning in the knowledge graph, and many newly proposed representation methods in the knowledge graph are often compared with it. The datasets applied to the methods in the ``Trans'' cluster can also be seen in Fig. \ref{fig:fig16}.

\begin{figure}[htbp]
\setlength{\belowcaptionskip}{-0.3cm}
    \centering
    \includegraphics[width=\linewidth]{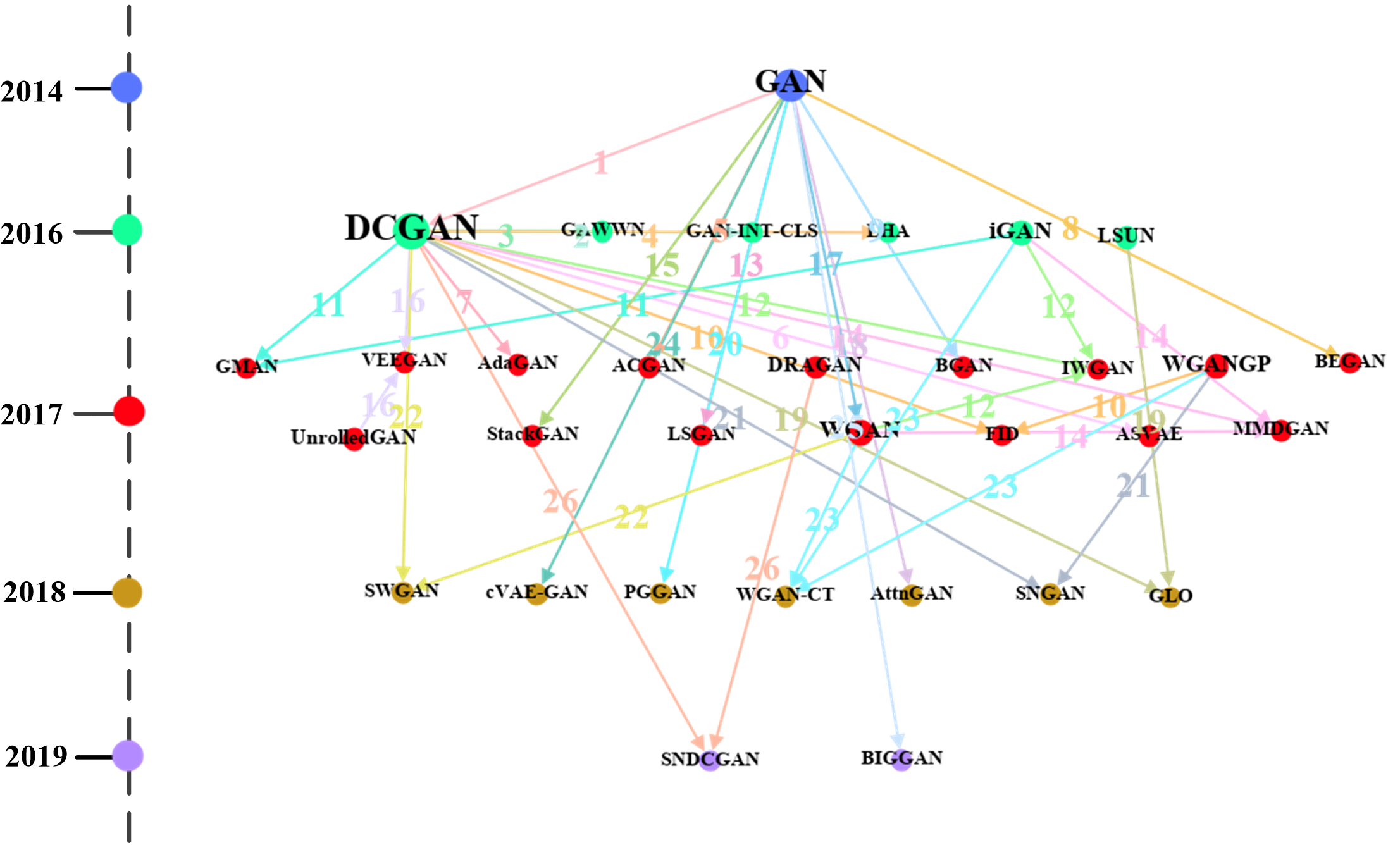}
    \caption{Roadmap of the methods in the "GAN" cluster. The numbers in the figure indicate the datasets used when comparing ${M_i}$ and ${M_j}$ in path ${M_i} \to {M_j}$. The coloured dots in the figure indicate the years, their sizes denote the size of the out-degrees, and the coloured lines signify the datasets represented by the numbers.\\The correspondence between the numbers and the datasets is as follows: 1: Face, NIST, SVHN and CelebA; 2: CUB(CU Bird), Oxford Flower and Oxford; 3: CUB(CU Bird), MPII Human, Caltech and MHP(Maximal Hyperclique Pattern); 4: ILSVRC and SVHN; 5: ImageNet; 6: NIST, CIFAR and ImageNet; 7: NIST; 8: CelebA; 9: NIST, CIFAR and SVHN; 10: BLUR, LSUN, SVHN, CIFAR, Noise, CelebA and LSUN Bedroom; 11: NIST, SVHN and CIFAR; 12: Google, LSUN and LSUN Bedroom; 13: Google; 14: NIST, LSUN, CIFAR, CelebA and LSUN Bedroom; 15: CUB(CU Bird) and Oxford; 16: NIST and CIFAR; 17: LSUN, CIFAR and LSUN Bedroom; 18: ImageNet and COCO; 19: NIST, SVHN, LSUN, CelebA and LSUN Bedroom; 20: LSUN, CelebA and LSUN Bedroom; 21: null; 22: NIST, LSUN, CIFAR, CelebA and LSUN Bedroom; 23: NIST, SVHN and CIFAR; 24: poem and Chinese poem; 25: CONFER; 26: null
}
    \label{fig:fig17}
\end{figure}

Figure \ref{fig:fig17} is the roadmap generated for the "GAN" cluster. After the content of the paper published by Hong et al. \cite{r31} is checked, the roadmap of the methods in our "GAN" cluster covers 75\% of the generative adversarial methods in \cite{r31}. Fig. \ref{fig:fig17} also shows the year when each method was proposed. For example, GAN and DCGAN were proposed in 2014 and 2016, respectively. Moreover, the out-degree of the DCGAN method is the largest. On the one hand, many methods, such as AdaGAN and SNDCGAN, are inspired by the DCGAN method and expand new methods. On the other hand, DCGAN is a representative method of generative adversarial networks, and many newly proposed generative adversarial methods are often compared with it. The datasets applied to the methods in the "GAN" cluster can also be seen in Fig. \ref{fig:fig17}.

\subsubsection{Results of research scene clusters}
The influencing degree rate of scene cluster $\backslash$s to scene cluster $s$ is obtained using Formula (1) in Section 3.6. The research scene clusters that are only influenced by one original paper or contain a few research scenes do not contain sufficient information. Moreover, the information of research scenes is messy in the research scene clusters that contain many research scenes. Accordingly, the research scene clusters that contain 15-20 research scenes (including 15 and 20) are analysed. The top three research scene clusters that are most easily affected by other research scene clusters are obtained; these clusters are color constancy, image memorability prediction and multiple kernel learning. The top three research scene clusters that are least likely to be affected by other research scene clusters are obtained; these clusters are significance detection, person re-identification and face recognition.

Using Formula (2) in Section 3.6, the method information with the largest influencing degree every year is obtained and shown in Table~\ref{tab:tab6}. Meanwhile, using Formula (3) in Section 3.6, the method information with the influencing degree rate every year is obtained and shown in Table~\ref{tab:tab7}.

The 12 methods in Table~\ref{tab:tab6} are mainly related to computer vision. In the methods with the largest influencing degree rate, 10 methods are related to computer vision. The methods in computer vision are likely to relatively affect other research scene clusters. 12 publication venues belong to Tier-A in CCF in Table~\ref{tab:tab6}. Meanwhile, 14 publication venues belong to Tier-A in CCF in Table~\ref{tab:tab7}. This finding indicates that the methods proposed by Tier-A publication venues easily influence other research scene clusters.

\begin{table}[htbp]
\caption{Method information with the largest influencing degree every year}
\label{tab:tab6}
\begin{tabular}{lcc}
\toprule
Method   & Year & Publication   Venue \\ \midrule
HOG      & 2005 & CVPR                \\
MILES    & 2006 & TPAMI               \\
ITML     & 2007 & ICML                \\
ESS      & 2008 & CVPR                \\
SRC      & 2009 & TPAMI               \\
LLC      & 2010 & CVPR                \\
RC       & 2011 & CVPR                \\
RootSIFT & 2012 & CVPR                \\
SDM      & 2013 & CVPR                \\
Glove    & 2014 & EMNLP               \\
FCN      & 2015 & CVPR                \\
ResNet   & 2016 & CVPR                \\
DenseNet & 2017 & CVPR                \\
ELMO     & 2018 & NAACL               \\
BERT     & 2019 & NAACL               \\ \bottomrule
\end{tabular}
\end{table}

\begin{table}[htbp]
\caption{Method information with the largest influencing degree rate every year}
\label{tab:tab7}
\begin{tabular}{lcc}
\toprule
Method  & Year & Publication Venue \\ \midrule
ASO     & 2005 & JMLR              \\
MILES   & 2006 & TPAMI             \\
SLDA    & 2007 & NeurIPS           \\
GPDM    & 2008 & TPAMI             \\
CDBN    & 2009 & ICML              \\
IMS     & 2010 & ACL               \\
FOCI    & 2011 & ICCV              \\
SOE     & 2012 & CVPR              \\
DECOLOR & 2013 & TPAMI             \\
CQ      & 2014 & ICML              \\
DALI    & 2015 & IJCV              \\
UHM     & 2016 & ACCV              \\
DDIS    & 2017 & CVPR              \\
LMN     & 2018 & AAAI              \\
ASTER   & 2019 & TPAMI             \\ \bottomrule
\end{tabular}
\end{table}

\section{Conclusions and Future Work}

Inspired by the idea of molecular marker tracing in the field of biochemistry, methods, datasets and metrics are used as markers for AI literature. The traces of these three named entities in the specific research process are used to study the development and change of the AI field. Firstly, the AI marker extraction model is used to extract AI markers from the "methodology" and "experiment" chapters in 122,446 AI papers. Effective methods and datasets are statistically analysed, and the annual development of the AI field is obtained. Secondly, we trace the original papers corresponding to effective methods and datasets. Statistical and propagation analyses are performed on the basis of the original paper tracing results. The results show that the number of effective methods proposed by Singapore, Israel and Switzerland is relatively large. The propagation of the effective methods on the datasets is rapidly increasing with the development of time. The effective methods proposed by China in recent years have an increasing influence on other countries, whilst France is the opposite. Finally, datasets and metrics are combined into AI research scenes. The methods and research scenes are clustered. The roadmaps of the methods are drawn on the basis of the method clusters and associated datasets to study the evolution of the methods in the same cluster. The influencing degree of the methods on the research scenes and that amongst research scenes are analysed on the basis of the clustering results of research scenes. The results show that saliency detection, a classic computer vision research scene, is the least likely to be affected by other research scenes.

In future work, we will improve the AI marker extraction model to optimise its extraction performance and try to extract AI markers from the tables and figures of AI literature to achieve a more comprehensive and accurate extraction of AI markers. The development of the AI field will also be efficiently demonstrated.

\bibliographystyle{ACM-Reference-Format}
\bibliography{main}

\appendix
\section{Normalisation strategies}
\subsection{Methods}

1) The digits in the methods are removed, except for some special methods, such as "C4.5" and "ID3". If the method is in the plural form, then it is converted to the singular form. For example, "SVMs" is normalised to "SVM".

2) After the digits are removed and converted to singular form, the methods with the same lowercase form are normalised to the same form.

3) After the words composed of lowercase letters are removed in the phrase, the methods with the same lowercase form are normalised to the same form.

4) The first letter of each word in the phrase are taken (if the word is composed of uppercase letters, then all the letters of the word are taken). Then, a query of whether a unique word exists in all methods corresponding to the phrase will be posted (that is, find the unique abbreviation corresponding to the full name). If a unique word exists, then the abbreviation and full name are normalised to "abbreviation (full name)". For example, "Long Short-Term Memory" and "LSTM" are normalised to "LSTM (Long Short-Term Memory)".

\subsection{Datasets}

1) The digits in the datasets are removed. If the dataset is in the plural form, then it is converted to the singular form. For example, "COLT 2011" is normalised to "COLT".

2) After the digits are removed and converted to singular form, the datasets with the same lowercase form are normalised to the same form.

3) If words whose first letter are capitalised are present in a phrase, then only the words whose first letter is capitalised are retained. For example, "Yale face" is normalised to "Yale".

4) The first letter of each word in the phrase is taken (if the word is composed of uppercase letters, then the letters of the word are taken). Then, a query of whether a unique word exists in all datasets corresponding to the phrase will be posted (that is, find the unique abbreviation corresponding to the full name). If a unique word exists, then the abbreviation and full name are normalised to "abbreviation (full name)".

\subsection{Metrics}

1) The digits in the metrics are removed. If the metric is in the plural form, then it is converted to the singular form. For example, "error rates" is normalised to "error rate".

2) After the digits are removed and converted to singular form, the metrics with the same lowercase form are normalised to the same form.

3) The metric is normalised into "recall", "accuracy", "precision", "speed" or "error rate", respectively, as long as the metric contains the following words: recall, accuracy, precision, speed or error rate. For example, the "mean accuracy" and "predictive accuracy" that contain "accuracy" are normalised to "accuracy".

4) The metric is normalised into "F-measure" as long as the metric contains the following words: F-score, F-measure, macroF, microF and F1.

5) If a word in a phrase is composed of uppercase letters, and the last word of the phrase is not rate, ratio or error, then only words composed of uppercase letters are retained. For example, "ACC information" is normalised to "ACC", and "RMS error" is normalised to "RMS error".

6) The first letter of each word in the phrase are taken (if the word is composed of uppercase letters, then all the letters of the word are taken). Then, a query of whether a unique word exists in all metrics corresponding to the phrase will be posted (that is, find the unique abbreviation corresponding to the full name). If a unique word exists, then the abbreviation and full name are normalised to "abbreviation (full name)".

\end{document}